\documentclass[useAMS,usenatbib,usegraphicx]{mn2e}
\usepackage{amsmath}
\usepackage[T1]{fontenc} 
\usepackage{times}
\usepackage{flushend} 
\usepackage{color}
\title[Narrow band SN Ia surveys]{Photometric type Ia supernova surveys in narrow band filters}
\author[H. S. Xavier et al.]
{Henrique S. Xavier$^{1,2}$\thanks{E-mail: hsxavier@if.usp.br}, L. Raul Abramo$^1$, Masao Sako$^2$, Narciso Ben\'{i}tez$^{3}$,  
\newauthor Maur\'{i}cio O. Calv\~{a}o$^{4}$, Alessandro Ederoclite$^{5}$, Antonio Mar\'{i}n-Franch$^{5}$,  
\newauthor Alberto Molino$^{3}$, Ribamar R. R. Reis$^{4}$, Beatriz B. Siffert$^{4}$ and Laerte Sodr\'{e} Jr.$^{6}$\\
$^{1}$ Instituto de F\'{i}sica, Universidade de S\~{a}o Paulo, Rua do Mat\~{a}o, Travessa R, 187, S\~{a}o Paulo, SP 05508-090, Brazil\\
$^{2}$ Department of Physics \& Astronomy, University of Pennsylvania, 209 South 33rd Street, Philadelphia, PA 19104, USA\\
$^{3}$ Instituto de Astrof\'{i}sica de Andaluc\'{i}a (CSIC), Glorieta de la astronom\'{i}a S/N, Granada 18008, Spain\\
$^{4}$ Instituto de F\'{i}sica, Universidade Federal do Rio de Janeiro, Av. Athos da Silveira Ramos, 149, Rio de Janeiro, RJ 21941-972, Brazil\\
$^{5}$ Centro de Estudios de F\'{i}sica del Cosmos de Arag\'{o}n (CEFCA), Plaza San Juan 1, Planta 2, 44001, Teruel, Spain\\
$^{6}$ Instituto de Astronomia, Geof\'{i}sica e Ci\^{e}ncias Atmosf\'{e}ricas, Universidade de S\~{a}o Paulo, Rua do Mat\~{a}o, 1226, S\~{a}o Paulo, SP 05508-090, Brazil}

\begin{document}

\maketitle

\begin{abstract}

We study the characteristics of a narrow band type Ia supernova survey through
simulations based on the upcoming Javalambre Physics of the accelerating universe 
Astrophysical Survey (J-PAS). This unique survey has the capabilities of obtaining
distances, redshifts, and the SN type from a single experiment thereby
circumventing the challenges faced by the resource-intensive spectroscopic
follow-up observations. We analyse the flux
measurements signal-to-noise ratio and bias, the supernova typing performance, the ability
to recover light curve parameters given by the SALT2 model, the photometric redshift precision
from type Ia supernova light curves and the effects of systematic errors on the data. We show
that such a survey is not only feasible but may yield large type Ia supernova samples (up to 250
supernovae at $z<0.5$ per month of search) with low core collapse contamination ($\sim 1.5$ per cent),
good precision on the SALT2 parameters (average $\sigma_{m_B}=0.063$, $\sigma_{x_1}=0.47$ and $\sigma_c=0.040$)
and on the distance modulus (average $\sigma_{\mu}=0.16$, assuming an intrinsic scatter $\sigma_{\mathrm{int}}=0.14$),
with identified systematic uncertainties $\sigma_{\mathrm{sys}}\la0.10\sigma_{\mathrm{stat}}$. Moreover,
the filters are narrow enough to detect most spectral features and obtain excellent photometric
redshift precision of $\sigma_z=0.005$, apart from $\sim$ 2 per cent of outliers. We also present
a few strategies for optimising the survey's outcome. Together with the detailed host galaxy
information, narrow band surveys can be very valuable for the study of supernova rates, spectral
feature relations, intrinsic colour variations and correlations between supernova and host galaxy
properties, all of which are important information for supernova cosmological applications. 
\end{abstract}


\section{Introduction}
\label{sec:Introduction}

Supernovae (SNe) and their relations with their surrounding environment have been an 
active field of study for decades. Their progenitors and explosion mechanisms are not 
fully known and understood, nor are all their possible variations, sub-classes and behaviours 
\citep{Hamuy00mn,Sullivan06mn,Leonard07mn,Xavier13mn}. 
On top of that, SNe play a key role in other scientific fields like chemical evolution of 
intra and intergalactic medium \citep{Wyse85mn,Zaritsky04mn,Scannapieco06mn}, star formation 
rate in galaxies \citep{Tsujimoto99mn,Yungelson00mn,Seo13mn}, 
energetics of the interstellar medium \citep{Chevalier77mn}, galaxy cluster density and 
temperature profiles \citep{Suginohara98mn,Voit01mn}, and on measurements of the cosmological expansion history 
of the universe \citep{Riess98mn,Perlmutter98mn,Conley11mn,Sullivan11mn}. Many of these subjects are 
interconnected, and a better understanding of one is likely to positively influence the other.

SN studies are made more difficult due to their rarity and their transient nature: 
SN rates are of order unity per galaxy per century and they are visible only for a couple of months \citep{Carroll96mn}. 
Fortunately, their cosmological importance have driven and continues to drive astrophysical 
surveys that can amass a relatively large number of such events. These surveys -- such as the Supernova 
Legacy Survey \citep[SNLS,][]{Pritchet05mn, Astier06mn}, the ESSENCE supernova survey 
\citep{Miknaitis07mn}, the Sloan Digital Sky Survey \citep[SDSS,][]{York00mn, Frieman08mn}, 
the Dark Energy Survey \citep[DES,][]{DES05mn,Bernstein12mn}, Pan-Starrs \citep{Kaiser10mn} 
and the Large Synoptic Survey Telescope \citep[LSST,][]{Ivezic08mn, LSST09mn} -- are broad 
band photometric surveys backed up by spectroscopic measurements. An appropriately 
time-distributed sequence of observations in a few broad band filters can provide a good 
measurement of the SN light curves up to high redshifts, while spectroscopy was indispensable 
for typing the SN and measuring its redshift.

Even though these projects could obtain images of a huge amount of SN candidates, their 
typing (a fundamental part in a SN program) was strongly based on their spectral 
features. While secure, this method is costly and time consuming, therefore it severely 
limits the SN sample sizes, specially since SN science must share time with different 
surveys goals. For instance, SDSS database contains near 660 spectroscopically 
confirmed SNe out of a total of $\sim 4650$ photometric SNe candidates (14 per cent), 
and DES expects to measure the spectra of 800 type Ia SNe from a total of 4000 SNe Ia with 
host galaxy spectroscopy (20 per cent) \citep{Bernstein12mn}. 
 
With this bottleneck in mind, a lot of effort was placed on photometrically typing SN 
candidates \citep[e.g.][]{Kessler10mn} and a lot of progress was achieved in this field 
\citep[e.g.][]{Sako11mn}. Even though typing can be reasonably good for SNe Ia without 
spectroscopy, a precise redshift prior is still needed in order to get good constraints 
on SN properties (specially colour), and this prior has to be obtained 
with spectroscopic measurements of the SN's host galaxy. This change in spectroscopy target 
(from the SNe to their hosts) facilitates the observations by allowing 
the measurements to be made well after the SNe have vanished, but it still presents a bottleneck 
for SN samples. From SDSS, $\sim 2500$ of the SNe without 
direct spectroscopic measurement also did not have spectroscopy from its host. These 
purely photometric SN samples will grow in the future as new surveys such 
as the LSST are expected to detect and measure the light curve of $\sim 10^7$ SNe 
\citep{LSST09mn}.
  
Spectroscopy is not only beneficial for SN light curve fitting and for measuring its redshift: it also 
conveys information about the SN properties. For instance, studies have indicated 
that SN Ia spectral features like the width of the SiII line and various flux ratios 
can be used to improve distance measurements 
\citep{Nugent95mn,Bongard06mn,Hachinger06mn,Bronder08mn,Foley08mn,Bailey09mn,Chotard11mn,Nordin11mn}. 
On top of that, SN Ia spectroscopy can help us to distinguish between various models 
for their luminosity intrinsic scatter \citep{Kessler13mn}. These measurements 
do not require high resolution spectra since the SN absorption features are reasonably 
large \citep[for a review, see][]{Filippenko97mn}.

SN science benefits also from spectroscopy of the SN host galaxies. An accurate 
measurement of the hosts properties -- or, even better, of the environment in the 
vicinity of the SNe -- will help to pin down their possible progenitors \citep{Galbany12mn}. 
Besides, the SN environment was shown to correlate with their rates and properties 
\citep{Sullivan06mn,Dilday10mn,Gupta11mn,Li11mn,Xavier13mn}, which are important 
information for stellar and chemical evolution of galaxies and galaxy clusters, and 
for cosmological distance measurements \citep{Sullivan10mn,Lampeitl10mn}. 

Given the importance of spectroscopic data and the challenges of obtaining it in 
large scale, we investigate the expected characteristics of a photometric SN Ia survey 
performed with a set of contiguous narrow band filters. Filters with transmission functions about 100--200 \AA\, 
wide still have enough resolution to detect almost every SN spectral feature. Since it acts 
as a low resolution spectrograph equipped with an integral field unit (IFU), all SNe 
detected by this type of survey automatically have their spectra measured. Moreover, it 
naturally yields rich information about their local environments.  

\begin{figure*}
\includegraphics[width=1\columnwidth]{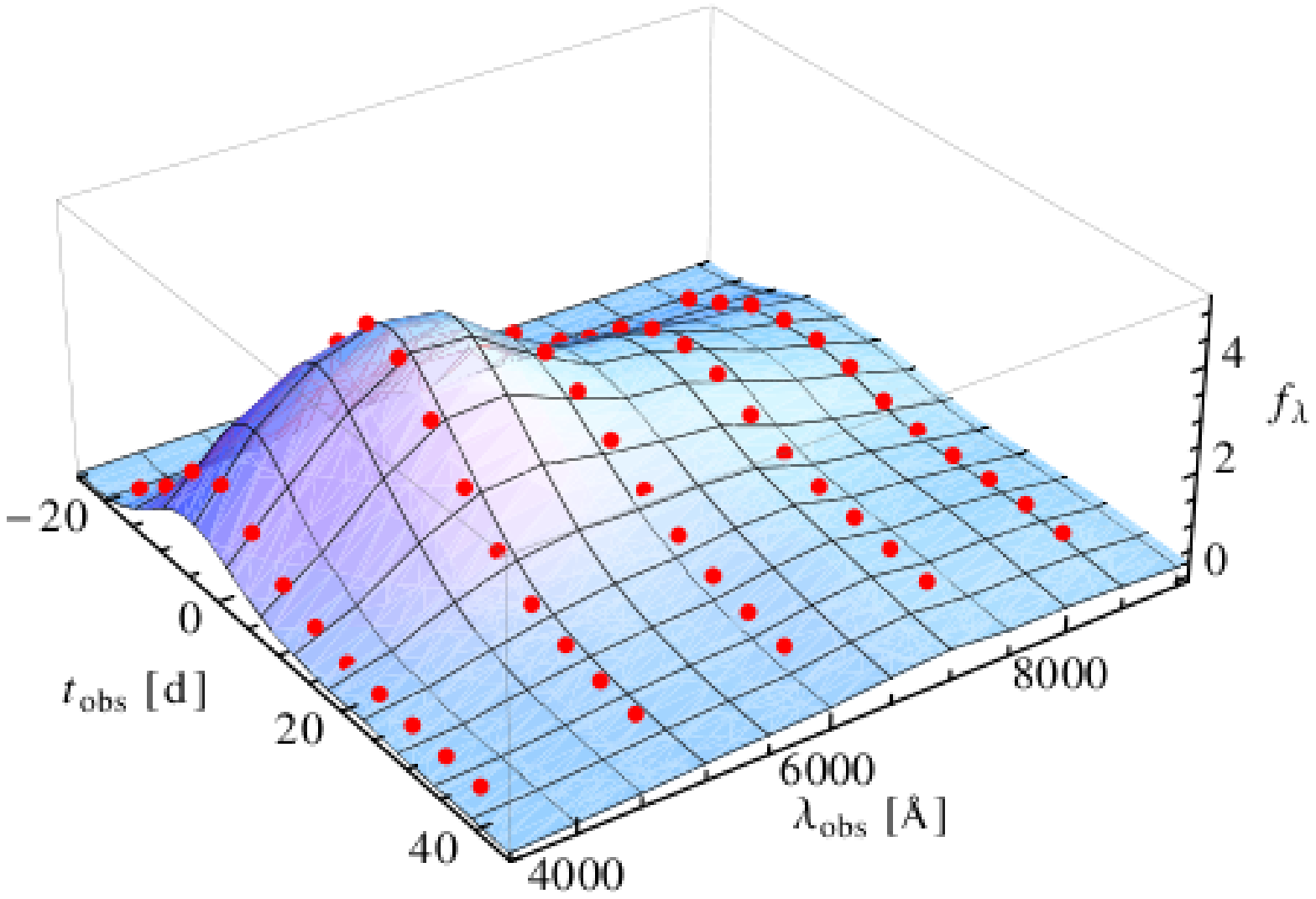}
\includegraphics[width=1\columnwidth]{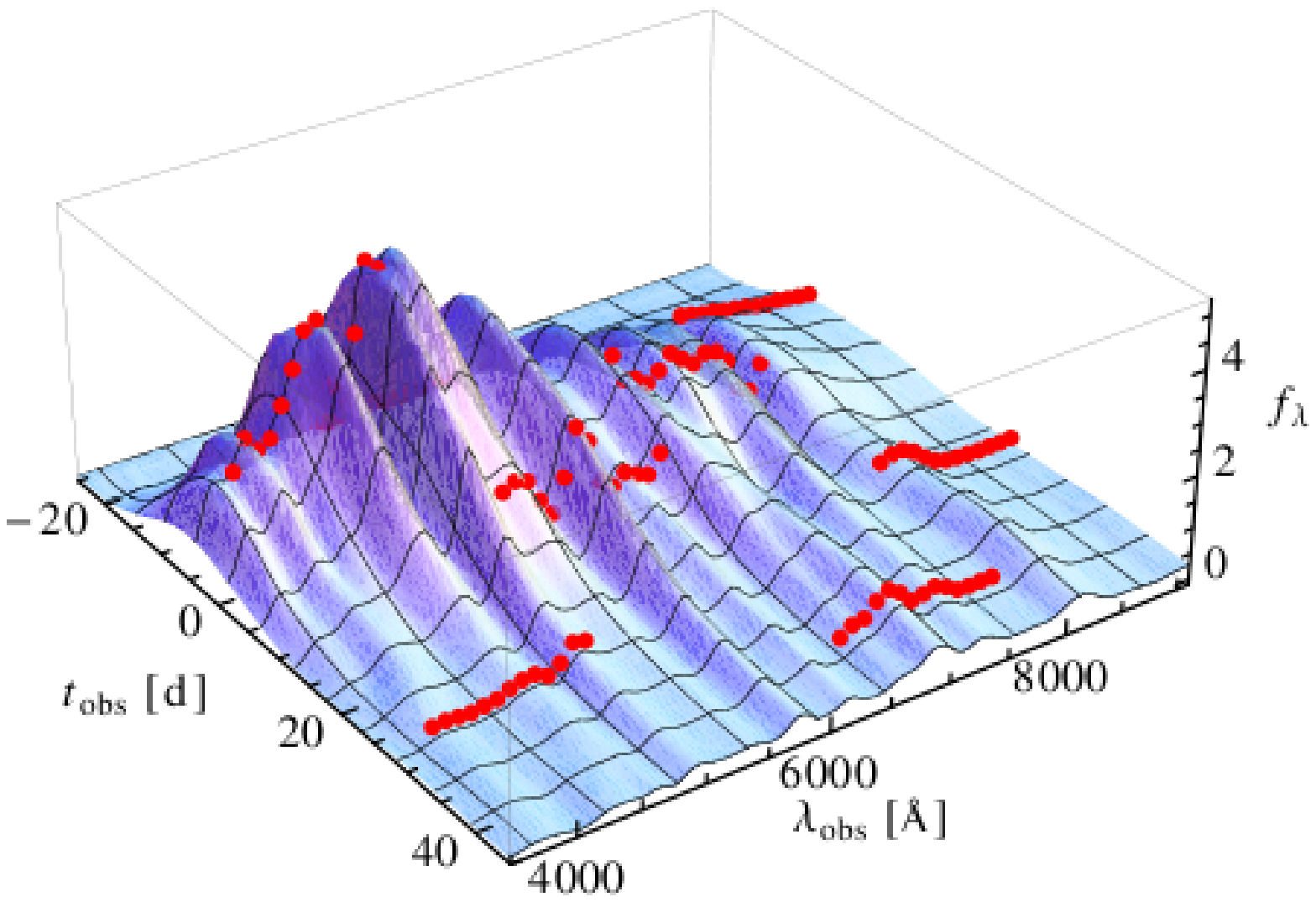}
\caption{SN Ia spectral surface at redshift $z=0.25$ convolved with top-hat functions $T(\lambda)$ 
1000 \AA\; (left panel) and 100 \AA\; wide (right panel), representing the resolution 
attainable with broad and narrow band filters, respectively. The epoch $t_{\mathrm{obs}}$ is given 
in days from maximum luminosity and the flux density $f_{\lambda}$ is in arbitrary 
units. SN observations can be represented as $N$ sampling points 
at specific epochs and filters central wavelengths $\{(\lambda_1,t_1),\ldots,(\lambda_N,t_N)\}$.
A typical broad band sampling and a possible narrow band sampling are shown in the left 
and right panels, respectively.  
Filters with bandwidths $\sim 100$ \AA\; can detect practically all SN spectral features 
while filters $\sim 1000$ \AA\; wide can only detect large scale features.}
\label{fig:spec-surface-res} 
\end{figure*}

The main features of a narrow band SN survey are best described with the help of 
Fig. \ref{fig:spec-surface-res}, which presents the spectral energy distribution of a 
SN Ia as a function of time $f_{\lambda}(\lambda_{\mathrm{obs}},t_{\mathrm{obs}})$ -- called spectral surface -- with 
typical wavelength resolutions for broad band (left panel) and narrow band filters (right panel). Measurements can be 
interpreted as sampling these surfaces at specific points, and the amount of information 
available for a narrow band survey is clear. For instance, while for a broad band survey 
the SN redshift must be inferred from the position of a wide peak, it can be 
inferred from the position of all the spectral peaks and troughs in the case of a narrow band survey. 
Fig. \ref{fig:spec-surface-res} also emphasises that the relevant SN quantities to 
be well sampled and constrained are \emph{not} individual light curves (spectral 
surface slices at fixed $\lambda$) but the entire collection of correlated 
light curves (i.e. the spectral surface itself). The individual light curve perspective 
is common in broad band SN surveys given they can only sample the spectral surface (Fig. 
\ref{fig:spec-surface-res}, left panel) at 1--5 different wavelengths. Since narrow 
band surveys may sample the spectral surface (Fig. \ref{fig:spec-surface-res}, right panel) 
at 20--60 different wavelengths, a large sampling in the same wavelength is not 
as important and one or two might suffice -- provided the observations in different 
wavelengths are also appropriately distributed in time.   

In this paper we make forecasts of SN Ia data obtainable with a narrow band filter 
survey by simulating and fitting light curves with the SALT2 model \citep{Guy07mn} as 
implemented by the {\sc snana} software package \citep{Kessler09amn}, using the 
Javalambre Physics of the accelerating universe Astrophysical Survey \citep[J-PAS,][]{Benitez14mn} 
as our fiducial survey. We estimate the performance of such a narrow band survey regarding 
the amount of observable SNe Ia, their average error and bias on various parameters, 
their redshift distribution and their typing purity and completeness, and compared 
with results for broad band surveys, namely SDSS and DES. Medium band 
surveys were already performed in the past [e.g. the COMBO-17 survey \citep{Wolf03mn} used 
12 filters about 220 \AA\, wide and the ALHAMBRA survey \citep{Moles08mn,Benitez09amn,Molino14mn} used 20 filters 
$\sim 310$ \AA\, wide], while narrow band surveys -- J-PAS\footnote{http://j-pas.org} 
and PAU\footnote{http://www.pausurvey.org} \citep{Benitez14mn, Marti14mn} -- are already being implemented. 
Besides, future spin-offs like a southern copy of J-PAS are under planning.

The outline of this paper is as follows: in section \ref{sec:simulation-inputs} 
we describe all the inputs we used to simulate the SN data, starting from 
light curve models and their allowed range of parameters (section \ref{sec:lightcurve-model}). 
Sections \ref{sec:fiducial-survey} and \ref{sec:survey-strategy} describe our fiducial 
survey, including its filter system and observing 
strategy. Host galaxy inputs and various noises estimates are described in sections 
\ref{sec:hostlib} and \ref{sec:noise-sources}. Our simulation results are 
presented in section \ref{sec:results}: the expected number of SNe per season 
and its redshift distribution; the flux measurement signal-to-noise ratio in each redshift 
and filter (section \ref{sec:flux-measurements}); the SN typing efficiency, 
SN Ia light curve parameters recovery and distance measurement precision 
(sections \ref{sec:typing-results} and \ref{sec:parameter-recovery}); and the 
quality of redshift inference from SN Ia light curves (section \ref{sec:sne-photoz}). 
A few suggestions for optimising a narrow band SN survey are presented in section 
\ref{sec:optimization}, and some systematic uncertainties are discussed in section 
\ref{sec:systematics}. Our conclusions and a summary of our main findings are 
presented in section \ref{sec:conclusions}.

\section{Simulation characteristics}
\label{sec:simulation-inputs}

{\sc Snana} can make realistic simulations of supernova surveys by generating 
different SN light curves at various redshifts according to a specified rate, 
then applying noise to the data -- based both on their intrinsic 
properties and on the survey apparatus -- and selecting the actually detected 
SNe based on defined selection cuts and on the survey design. The simulated data 
can then be typed and fitted just like real data.

To perform the simulations, the following inputs are required: a SN light curve model and 
distributions for its parameters; a SN rate as a function of redshift; a library of 
potential host galaxies, used to introduce extra noise and to possibly supply a 
redshift prior to the SNe; either the SN position or the value of the Milky Way 
excess colour in order to calculate the Galactic extinction; the filters transmission 
functions; an observation schedule listing the days and filters used, along with the 
photometric conditions (zero points, sky noise, CCD readout noise and point spread 
function); the area covered by the survey; and eventual selection cuts that can 
be applied to the data. These are described in detail below. We also briefly describe 
the typing and fitting methods used by the {\sc snana} package.
      
\subsection{Light curve models}
\label{sec:lightcurve-model}

\begin{table}
\caption{Light curves and spectra used as templates for simulating CC-SNe. 
The columns present, from left to right: a template identification; its type; the fraction 
of the simulated CC-SN light curves generated using each template; the mean absolute 
magnitude at peak in \emph{B} band, in the supernova rest-frame, and the standard deviation 
$\sigma_{M_B}$ of the light curves simulated with each template. These are the default 
{\sc snana} templates; all values (Frac., $M_B$ and $\sigma_{M_B}$) were set to match 
those reported in \citet{Li11mn}.} 
\label{tab:cc-sn-templates}
\centering
\begin{tabular}{|l|c|c|c|c|}
\hline
Template ID & Type &  Frac.  & $M_B$  & $\sigma_{M_B}$\tabularnewline\hline
SDSS-000018 & IIP & 0.0246 & -17.11 & 1.050 \tabularnewline
SDSS-003818 & IIP & 0.0246 & -15.09 & 1.050 \tabularnewline
SDSS-013376 & IIP & 0.0246 & -15.46 & 1.050 \tabularnewline
SDSS-014450 & IIP & 0.0246 & -16.16 & 1.050 \tabularnewline
SDSS-014599 & IIP & 0.0246 & -16.06 & 1.050 \tabularnewline
SDSS-015031 & IIP & 0.0246 & -15.25 & 1.050 \tabularnewline
SDSS-015320 & IIP & 0.0246 & -15.61 & 1.050 \tabularnewline
SDSS-015339 & IIP & 0.0246 & -16.32 & 1.050 \tabularnewline
SDSS-017564 & IIP & 0.0246 & -17.01 & 1.050 \tabularnewline
SDSS-017862 & IIP & 0.0246 & -15.68 & 1.050 \tabularnewline
SDSS-018109 & IIP & 0.0246 & -16.02 & 1.050 \tabularnewline
SDSS-018297 & IIP & 0.0246 & -15.28 & 1.050 \tabularnewline
SDSS-018408 & IIP & 0.0246 & -15.29 & 1.050 \tabularnewline
SDSS-018441 & IIP & 0.0246 & -15.37 & 1.050 \tabularnewline
SDSS-018457 & IIP & 0.0246 & -15.65 & 1.050 \tabularnewline
SDSS-018590 & IIP & 0.0246 & -14.52 & 1.050 \tabularnewline
SDSS-018596 & IIP & 0.0246 & -15.80 & 1.050 \tabularnewline
SDSS-018700 & IIP & 0.0246 & -14.32 & 1.050 \tabularnewline
SDSS-018713 & IIP & 0.0246 & -15.25 & 1.050 \tabularnewline
SDSS-018734 & IIP & 0.0246 & -14.84 & 1.050 \tabularnewline
SDSS-018793 & IIP & 0.0246 & -16.42 & 1.050 \tabularnewline
SDSS-018834 & IIP & 0.0246 & -15.71 & 1.050 \tabularnewline
SDSS-018892 & IIP & 0.0246 & -15.44 & 1.050 \tabularnewline
SDSS-020038 & IIP & 0.0246 & -17.04 & 1.050 \tabularnewline
SDSS-012842 & IIn & 0.0200 & -17.33 & 1.500 \tabularnewline
SDSS-013449 & IIn & 0.0200 & -16.60 & 1.500 \tabularnewline
Nugent+Scolnic & IIL & 0.0800 & -16.75 & 0.640 \tabularnewline
CSP-2004gv & Ib & 0.0200 & -16.69 & 0.000 \tabularnewline
CSP-2006ep & Ib & 0.0200 & -18.51 & 0.000 \tabularnewline
CSP-2007Y & Ib & 0.0200 & -15.53 & 0.000 \tabularnewline
SDSS-000020 & Ib & 0.0200 & -16.14 & 0.000 \tabularnewline
SDSS-002744 & Ib & 0.0200 & -15.99 & 0.000 \tabularnewline
SDSS-014492 & Ib & 0.0200 & -16.93 & 0.000 \tabularnewline
SDSS-019323 & Ib & 0.0200 & -16.11 & 0.000 \tabularnewline
SNLS-04D1la & Ibc & 0.0167 & -15.84 & 1.100 \tabularnewline
SNLS-04D4jv & Ic & 0.0167 & -14.50 & 1.100 \tabularnewline
CSP-2004fe & Ic & 0.0167 & -15.66 & 1.100 \tabularnewline
CSP-2004gq & Ic & 0.0167 & -14.93 & 1.100 \tabularnewline
SDSS-004012 & Ic & 0.0167 & -15.84 & 1.100 \tabularnewline
SDSS-013195 & Ic & 0.0167 & -15.63 & 1.100 \tabularnewline
SDSS-014475 & Ic & 0.0167 & -16.53 & 1.100 \tabularnewline
SDSS-015475 & Ic & 0.0167 & -14.43 & 1.100 \tabularnewline
SDSS-017548 & Ic & 0.0167 & -16.34 & 1.100 \tabularnewline
\hline
\end{tabular}
\end{table}

For simulating Core Collapse SNe (CC-SNe) we used the spectral templates 
available in the {\sc snana} package, listed in Table \ref{tab:cc-sn-templates}, 
which were based on objects observed by various surveys. Table 
\ref{tab:cc-sn-templates} also shows the fraction of simulated CC-SNe 
that was drawn from each template, the absolute AB magnitude in 
the \emph{B} band, in the supernova rest-frame, used to normalise it and a 
coherent (same for all epochs and wavelengths) random Gaussian deviation applied to 
the magnitude in each simulation of that template. These values were based on 
the work of \citet{Li11mn}. 
The extinction caused by host galaxy dust is modelled with the curve from 
\citet{ODonnell94mn}, a fixed ratio of total to selective extinction $R_V=2.22$ 
and an extinction at band \emph{V}, $A_V$, drawn from a distribution 
$f(A_V)=\exp(-A_V/0.334)$, limited to values $-1<A_V<1$.

For simulating and fitting SN Ia light curves, we used the SALT2 model which is 
adequate for narrow band filters since it returns sufficiently high resolution 
($\sim60$ \AA ) spectra for each epoch \citep{Guy07mn}. Since a narrow band survey 
is likely to detect more variation in the light curves and spectra than current 
models can predict, these are to be understood as general guides to how well such 
surveys can perform. 

SALT2 is an observer frame spectral model based on five parameters: the redshift $z$, a 
time of maximum luminosity $t_0$, a colour term $c$, a principal component factor  
$x_1$ which can be roughly interpreted as a stretch parameter, and an overall 
normalisation $x_0$, which can be translated into an apparent magnitude $m_B$ 
at peak in the SN rest-frame $B$ band. The observed spectral flux density $f_{\lambda}$ 
for a given epoch $t_{\mathrm{obs}}$ and wavelength $\lambda_{\mathrm{obs}}$ is given by:

\begin{equation}
f_{\lambda}(\lambda_{\mathrm{obs}},t_{\mathrm{obs}}) = \frac{x_0}{1+z} \left[ 
M_0 \left( \lambda, t \right) + x_1 M_1 \left( \lambda, t \right) \right] 
e^{c C\left( \lambda \right) }\; ,
\label{eq:salt2-flux}
\end{equation}  
where $\lambda$ and $t$ are the rest-frame wavelength and time from maximum, 
given by: $\lambda = \lambda_{\mathrm{obs}} / (1+z)$ and $t = (t_{\mathrm{obs}}-t_0)/(1+z)$. 
$M_0(\lambda, t)$ is a rest-frame average spectral surface (it gives you the average spectrum 
for each epoch $t$); $M_1(\lambda, t)$ is a principal component that accounts for the 
main deviations from $M_0$; and $C(\lambda)$ is a time-independent colour law that accounts 
for both intrinsic colour variations and dust extinction by the host galaxy. 

For each simulated SN, the redshift $z$ is randomly drawn according to the survey 
volume at each $z$ slice and to the CC-SN \citep{Kessler10mn} and SN Ia 
\citep{Dilday08mn} rates below:

\begin{equation}
\frac{\mathrm{d}N_{CC}}{\mathrm{d}z}=6.8\times10^{-5}(1+z)^{3.6}h_{70}^3\;\mathrm{Mpc^{-3}yr^{-1}}\;,
\label{eq:ccsn-rate}
\end{equation}

\begin{equation}
\frac{\mathrm{d}N_{Ia}}{\mathrm{d}z}=2.6\times10^{-5}(1+z)^{1.5}h_{70}^3\;\mathrm{Mpc^{-3}yr^{-1}}\;,
\label{eq:snia-rate}
\end{equation}
where $h_{70}=H_0/(70\mathrm{\;km\;s^{-1}\;Mpc^{-1}})$ and $H_0$ is the Hubble constant. 
The $x_1$ and $c$ parameters are drawn from Gaussian distributions with zero mean 
and standard deviations of 1.3 and 0.1, respectively, but constrained to the range 
$-5<x_1<5$ and $-0.4<c<0.4$. The time of maximum $t_0$ is drawn from a uniform 
distribution, and $x_0$ is calculated from the formula:

\begin{equation}
x_0 = 10^{-0.4(m_B-M-30)}\;,
\label{eq:x0-def}
\end{equation} 
\begin{equation}
m_B-M = \mu-\alpha x_1 +\beta c\;,
\label{eq:mB-def}
\end{equation} 
where $M$ is an average absolute magnitude, $\alpha$ and $\beta$ are positive 
constants that account for the fact that SNe Ia with broader light curves 
($x_1>0$) are usually brighter while redder SNe Ia ($c>0$) are usually dimmer. 
When simulating SNe Ia, these three quantities were fixed to $M=-19.365$, 
$\alpha=0.11$ and $\beta=2.60$ 
\citep{Richardson02mn,Kessler09bmn}\footnote{SALT2 magnitudes have an offset 
from AB magnitudes. The value of $M$ used in the simulations corresponds to 
an AB absolute magnitude $M_B=-19.095$ for $H_0=70\mathrm{\;km\;s^{-1}\;Mpc^{-1}}$.}. 
The distance modulus is defined as 
$\mu\equiv 5\log_{10}(\frac{d_\mathrm{L}}{10\mathrm{pc}})$, where $d_\mathrm{L}$ is 
the luminosity distance to the SNe Ia. To calculate $d_{\mathrm{L}}$ and the 
survey volume we assumed a flat $\mathrm{\Lambda CDM}$ 
cosmological model with $H_0=70\mathrm{\;km\; s^{-1}\; Mpc^{-1}}$ and $\Omega_\mathrm{m}=0.3$.
To simulate the SN Ia intrinsic scatter $\sigma_{\mathrm{int}}$ in Hubble diagrams, 
we introduced a 0.14 mag scatter in the $m_B$ calculated from Eq. \ref{eq:mB-def}.

\subsection{Our fiducial survey}
\label{sec:fiducial-survey}

\begin{figure*}
\includegraphics[width=1\textwidth]{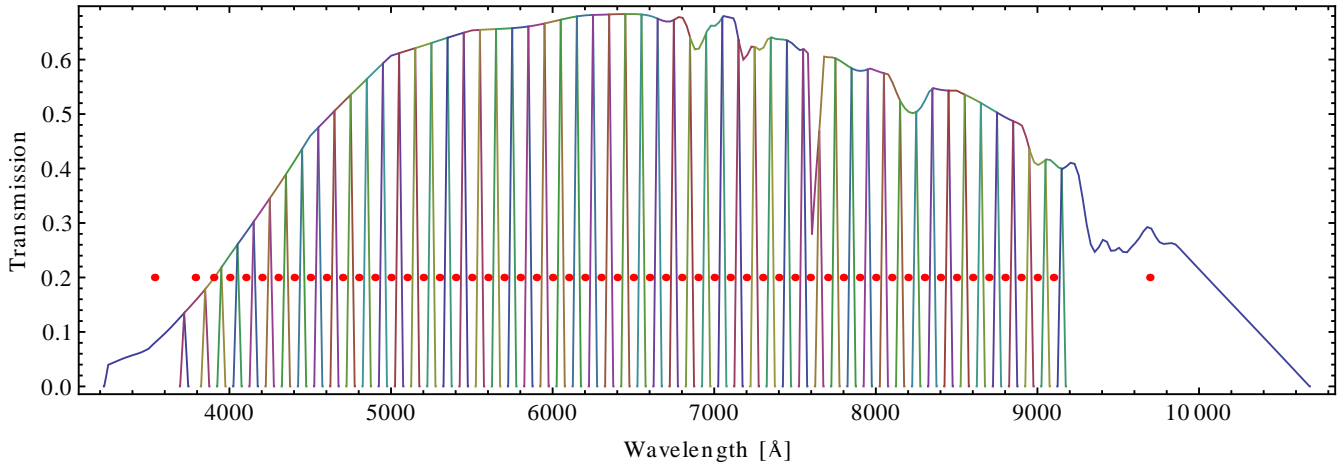}
\caption{Main 56 J-PAS filters transmission functions used in this work. The 
red dots indicate the central wavelength $\lambda_{\mathrm{c}}$ position of each 
filter (the position in the vertical axis is arbitrary). The transmission 
curves already include atmosphere transparency and CCD and optics efficiency.}
\label{fig:filter-transmission} 
\end{figure*}

We based the inputs needed for our simulations on the J-PAS survey. J-PAS 
is an 8500 $\mathrm{deg}^2$ survey aimed at measuring the baryon acoustic 
oscillations (BAO) at various redshifts using a few broad band (\emph{ugr} 
filters plus two unique filters) and 54 narrow band optical filters. It is 
expected to start taking data in 2015 using a newly built, large field of 
view (7 $\mathrm{deg}^2$ at full focal plane coverage), dedicated 2.5-m 
telescope situated at Sierra de Javalambre, in mainland Spain, equipped with 
a 14 CCD camera covering 67 per cent of the focal plane. The J-PAS is described in detail in 
\citet{Benitez14mn} and is an updated version of the survey described 
in \citet{Benitez09mn}. By using an existing project as our fiducial survey, 
we force our simulations to stay within more realistic boundaries.

For this study we used the main contiguous J-PAS filters described in Fig. 
\ref{fig:filter-transmission}. These are 54 narrow band filters with width 
$\sim 145$ \AA\, and spaced by 100 \AA, plus two broader filters at the ends of 
the wavelength range. For convenience, we will number them from 1 to 56 following 
their order in central wavelength $\lambda_{\mathrm{c}}$ (e.g. the bluest filter is number 1, the 
reddest filter is 56 and the reddest narrow band filter is 55). Each individual 
exposure, in each filter, will be of 60 s for filters 1--42 (3500 \AA\ 
$\leq\lambda_{\mathrm{c}}\leq$ 7800 \AA) and 120 s for filters 43--56 (7900 \AA\ 
$\leq\lambda_{\mathrm{c}}\leq$ 9700 \AA). 

For simulating photometric data, other characteristics of the telescope, 
camera and site are necessary. We assumed an effective aperture of 223 cm, 
a plate scale of 22.67 $\mathrm{arcsec/mm}$ and a pixel size of 10 
$\mathrm{\mu m}$. The CCD readout noise was set to 6 electrons per pixel, 
with a readout time of 12 s. These values are very close to the ones reported 
for J-PAS \citep{Benitez14mn}. The point spread function was modelled as a Gaussian 
with dispersion $\sigma$ determined by a conservative estimate of the seeing 
(0.8 arcsec) at the J-PAS site \citep[Observatorio Astrof\'{i}sico de Javalambre, OAJ,][]{Moles10mn}. 

\begin{figure}
\includegraphics[width=1\columnwidth]{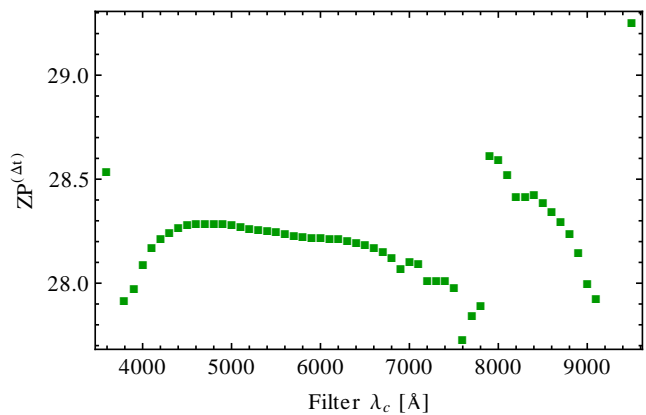}
\caption{Average zero points for the 56 filters. The large difference 
from filters 1 to 2 and 55 to 56 are due to their band size, and the 
difference between filters 42 and 43 (around 7900 \AA) is due to the 
doubling in exposure time.}
\label{fig:zero-points} 
\end{figure}

With this information, we can calculate the zero point that relates an object's 
magnitude to its corresponding CCD electron count, an {\sc snana} required input. 
Note that the zero points we refer to in this paper relate magnitudes 
to total counts instead of count rates as this is {\sc snana}'s convention; 
to stress this fact we will call them $ZP^{\mathrm{(\Delta t)}}$. 
By assuming that the object's spectrum is fairly constant within a filter wavelength 
range and using the AB magnitude system, we calculated the zero points $ZP_n^{\mathrm{(\Delta t)}}$ for 
the filter $n$ as:

\begin{equation}
ZP_n^{\mathrm{(\Delta t)}} = 2.5 \log_{10} \left[ \frac{\pi D^2\Delta t_n}{4h}\int\frac{T_n(\lambda)}{\lambda}d\lambda \mathrm{\frac{erg}{cm^2}}\right] -48.6\;, 
\label{eq:zero-point}
\end{equation}   
where $D$ is the telescope aperture, $h$ is the Planck's constant, $\Delta t_n$
and $T_n(\lambda)$ are the filter $n$ exposure time and transmission function, 
respectively. Fig. \ref{fig:zero-points} shows the average zero 
points used in the simulations.   

When comparing narrow band to broad band surveys, more attention will 
be given to SDSS rather than DES since the former is much more similar to our 
fiducial survey: it also used a 2.5-m telescope with a field of view (FoV) 
of 7 $\mathrm{deg^2}$, and its exposure time was 55 s \citep{Gunn06mn,Frieman08mn}. 
In contrast, DES uses a 4-m telescope with a 3 $\mathrm{deg^2}$ FoV and an average 
exposure time of 230 s for its shallow fields \citep{Bernstein12mn}.

\subsection{Survey strategy}
\label{sec:survey-strategy}

Most astrophysical surveys have multiple goals and the final survey strategy 
may be a compromise between optimal strategies for different sciences. 
However, the SNe need for a particular observation schedule usually excludes 
them from the main parts of surveys. For instance, the SDSS supernova survey 
was restricted to the months of September through November, during the years 
2005--2007, scanning a region of $\sim 300$ $\mathrm{deg^2}$ \citep{Frieman08mn}. 
DES is expected to employ  $\sim 32$ per cent of its total time and $\sim 10$ per cent 
of its photometric time for SN science, imaging an area of 30 $\mathrm{deg^2}$ 
\citep{Bernstein12mn}. To test a possible optimisation of the survey's time usage, 
we analysed a strategy suitable both for SN and for a galaxy survey 
at the same time. This multi-purpose strategy is termed 2+(1+1) and is likely to be 
adopted by J-PAS (although its particular implementation might not involve 
all its filters during the same observing season). 

\begin{figure}
\includegraphics[width=1\columnwidth]{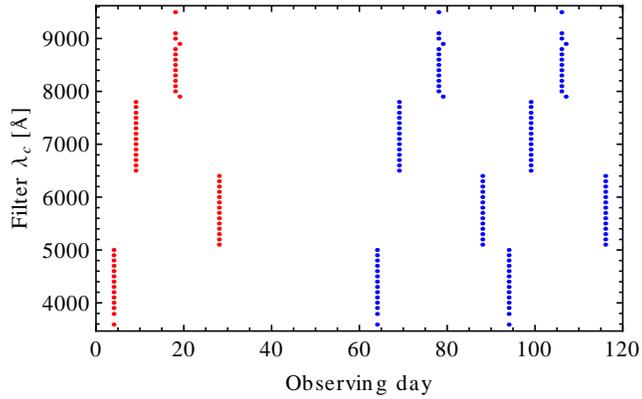}
\caption{Example of observation schedule for our main scenario. The red and blue 
points represent the template and search observations, respectively. The filters being 
used in each day are identified by their central wavelengths $\lambda_{\mathrm{c}}$. 
Provided there is a gap of at least one month between template and search, the exact template 
schedule is irrelevant for SNe. Schedule realisations are affected by weather conditions 
and the SN angular position.}
\label{fig:cadence-M01} 
\end{figure}

The 2+(1+1) strategy consists of a total of four exposures per filter per field. 
From the SN science perspective, the first two (which we call `template') are 
used to form an image of the SN environment and are taken on the same night. 
If a SN shows up in the last two exposures (which we call `search'), the template 
is used to subtract the host galaxy. Therefore, it is important to leave a minimum 
time gap of approximately one month between the last template observation and the 
first search observation so the templates are not contaminated by the SNe. Fig. 
\ref{fig:cadence-M01} presents a typical observation schedule for a given field.

In a real survey, the two template exposures would be combined into a deeper image; 
in this case, the errors on the two search exposures taken in the same filter would 
be correlated after the template subtraction, and this correlation would have 
to be taken into account during the data analysis. Unfortunately, 
{\sc snana}\footnote{The {\sc snana} version used was {\sc v10\_29.}} was originally designed for surveys with very 
deep template images and does not fully support such analysis. To get around this 
issue we did not combine the two template exposures and used each one to subtract from a 
different search exposure. This approach trades the correlation between measurements for 
higher noise in each observation, thus making our simulations a conservative estimate of 
the capabilities of the survey.

During the search observations each field is imaged in 8 different epochs 
separated by $\sim 1$ week, and in each epoch the field is imaged using 14 
different filters, making a total of 112 search observations (twice in each filter). 
The two exposures taken in the same filter are separated 
by $\sim 1$ month and it takes around two months to complete all search observations. 
In our main scenario schedule, 
the 14 filters that are observed in the same day are contiguous, which 
allows for the imaging of specific parts of the SN spectrum. Variations on this 
scenario are presented in section \ref{sec:optimization}.
In our simulations, the SN observation schedule might be altered by the 
SN position in the field and by weather conditions which introduce a 0.16 
chance of delaying a measurement. 
Moreover, complex particularities of the J-PAS filter positioning on the 
focal plane -- whose specifications are beyond the scope of this paper -- 
increase the survey's footprint at expense of full filter coverage in 
some regions. In our fiducial SN survey, this translates into a 0.24 chance 
of 8 or more filters not being observed at all and into an effective FoV of 5.4 
$\mathrm{deg^2}$. Assuming 8 hours of night time per day and the exposure 
and readout time from section \ref{sec:fiducial-survey}, this strategy 
can be applied to $\sim 800$ $\mathrm{deg^2}$.

\subsection{Host galaxy library}
\label{sec:hostlib}

Our simulations made use of a library of host galaxies for two purposes: 
introducing extra Poisson noise left over after the host galaxy subtraction 
and for supplying a redshift prior when fitting the SN light curves. 
For each entry, the library contained the galaxy's true redshift, its angular 
major and minor axis at half light (we used deVaucouleurs profiles), an orientation 
angle, the observed magnitude in each of the survey's filters, and its photometric 
redshift (photo-$z$) and corresponding error. The orientation angle was drawn from an 
uniform distribution, while the luminosity profiles and magnitudes were drawn 
from actual SDSS data \citep{Abazajian09mn} for SN host galaxies \citep{Gupta11mn}. 
To compute the magnitude of the host galaxies in the J-PAS filters we fitted SDSS DR5 spectral 
templates\footnote{http://www.sdss.org/dr5/algorithms/spectemplates/} to 
SDSS broad band photometry and used the best-fitting spectrum to generate the 
narrow band fluxes.
 
The luminosity profile, the orientation angle and the observed magnitudes 
are only used to generate the extra Poisson noise in the SN photometry. A random 
galaxy at a similar redshift of the SN is chosen, along with the SN's 
position on the galaxy, and then the flux coming from the galaxy is calculated. 
This flux is used to compute a CCD count which in turn serve as the mean value 
for a Poisson distribution from which the galaxy noise is drawn. 
Since the process of image subtraction increases the noise coming from the 
host galaxy (photon counts from the galaxy may vary between different exposures), we 
emulated this noise increase by making the galaxies brighter 
by a factor of $(1+1/N)$ [their magnitudes were decreased by $-2.5\log_{10}(1+\frac{1}{N})$, 
see Appendix \ref{sec:snr-model}], 
where $N$ is the number of images of the galaxy alone, taken with 
the same exposure time as the SNe, that are combined into a single subtraction 
image. 
In our simulations, $N=1$. In any case, the host galaxy Poisson noise contribution proved to be sub-dominant 
when compared to other sources of errors (see section \ref{sec:flux-measurements}).

Although real galaxy photo-$z$s are usually non-Gaussian, we adopted Gaussian 
errors for simplicity\footnote{Moreover, the {\sc snana} version 10\_29 used 
in this work does not support different distributions.}. The photo-$z$s and 
corresponding errors adopted in our host galaxy library were based on J-PAS expected 
precision, which \citet{Benitez09mn,Benitez14mn} 
reported to be $0.003(1+z)$ for luminous red galaxies. Since not all galaxies 
may reach this error level we used a fixed precision of $0.005(1+z)$ for all 
SN hosts. This level of galaxy photo-$z$ accuracy is unique to narrow band surveys 
and, as shown is section \ref{sec:sne-photoz}, can also be attained from the SNe 
themselves.  

\subsection{Noise sources}
\label{sec:noise-sources}

The {\sc snana} software includes many sources of noise to the simulated 
photometry: Poisson fluctuations from the source, the host galaxy and from the sky; 
CCD readout noise; a multiplicative flux error; and error on the Milky Way extinction 
correction. The multiplicative error on the flux (which can be translated into an additive 
error on the magnitudes) models a combination of errors such as standard 
stars measurement, flat fielding and other photometric technique errors 
\citep{Smith02mn,Astier13mn,Kessler13mn}, and will be generically called 
calibration rms error $\sigma_{ZP}$. It is worth reminding that assuming a random 
error to model these effects is a simplification given that they might be 
correlated between different measurements. The CCD readout and the calibration rms 
errors were set to 6 electrons per pixel and 0.04 mag, respectively, for all 
filters, both conservative estimates. These and other relevant simulation 
parameters are summarised in Table \ref{tab:survey-facts}. The effects of systematic 
uncertainties on the data are analysed in section \ref{sec:systematics}.

\begin{figure}
\includegraphics[width=1\columnwidth]{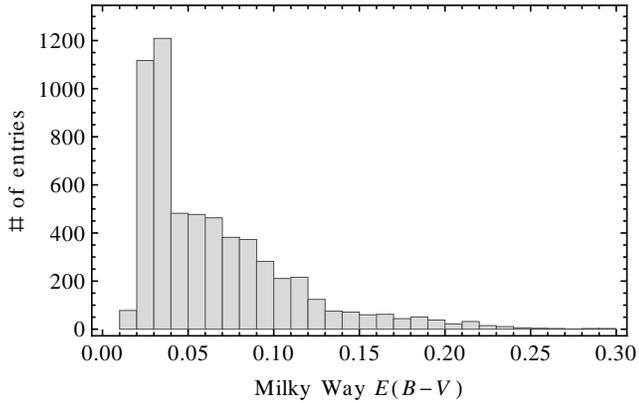}
\caption{Histogram of the Milky Way excess colour $E(B-V)$ sample used to 
compute Galactic extinction on the SN fluxes.}
\label{fig:mwebv} 
\end{figure}

The Galactic extinction correction error used was the {\sc snana} default (16 per cent). 
The true values of the excess colour $E(B-V)$ used for each SN simulation were 
drawn from a SDSS stripe 82 $E(B-V)$ sample presented in Fig. \ref{fig:mwebv}, and 
the extinction at each wavelength was calculated using the \citet{ODonnell94mn} 
curve and $R_V=3.1$.

\begin{figure}
\includegraphics[width=1\columnwidth]{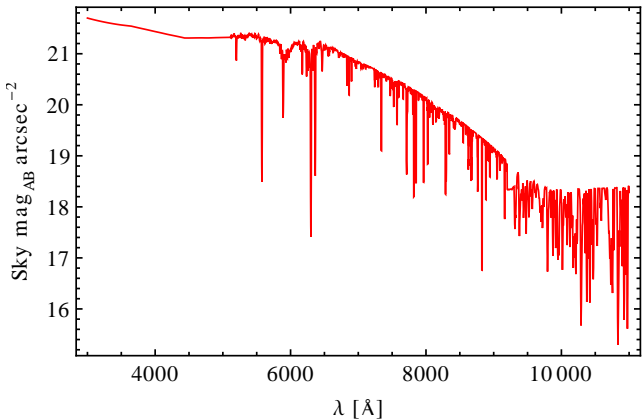}
\caption{Estimate of the average night sky spectrum for the J-PAS site, 
in AB magnitudes per $\mathrm{arcsec^2}$.}
\label{fig:sky-spectrum} 
\end{figure}

\begin{figure}
\includegraphics[width=1\columnwidth]{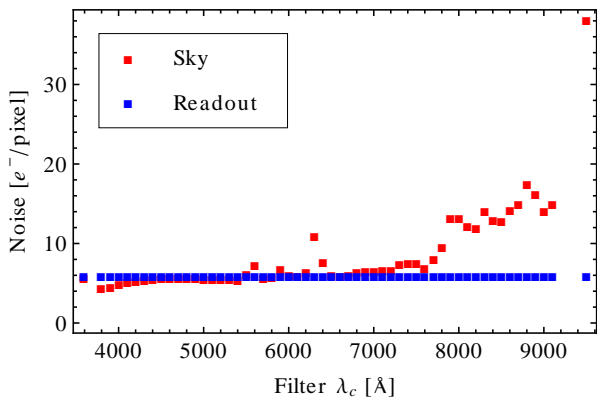}
\caption{Estimated sky (solid red line) and readout (dashed blue line) noise per pixel 
for the 56 filters. Due to the small bandwidth of most filters, both noises are comparable.}
\label{fig:sky-read-noise} 
\end{figure}

Based on an updated version of the sky spectrum from \citet{Benitez09mn}, 
presented in Fig. \ref{fig:sky-spectrum}, we estimated the photometry sky 
noise per pixel $\sigma_{\mathrm{sky},n}$ for filter $n$ using the following equation:  
 
\begin{equation}
\sigma_{\mathrm{sky},n}^2=\frac{\pi D^2}{4}\Delta t_n P^2\int f_{\mathrm{sky},\lambda}(\lambda)T_n(\lambda)\frac{\lambda}{hc}d\lambda\;,
\label{eq:sky-sig}
\end{equation}
where $P$ is the pixel angular size in arcsec, $c$ is the speed 
of light and $f_{\mathrm{sky},\lambda}$ is the sky spectral energy density (SED) per 
$\mathrm{arcsec^2}$. Fig. \ref{fig:sky-read-noise} presents the obtained 
$\sigma_{\mathrm{sky},n}$ values and compare them with the CCD readout noise. 
Due to the narrow band nature of most filters, the adopted exposure time of 
60 s makes the sky noise and the readout noise comparable.

The process of image subtraction required in SN surveys for removing the 
host galaxy flux increases the final SN photometry error. To account for 
this fact, we introduced this extra noise in the sky noise budget. The 
actual sky noise used in the simulations $\sigma'_{\mathrm{sky},n}$ is 
related to the pure sky noise $\sigma_{\mathrm{sky},n}$ by:

\begin{equation}
\sigma'_{\mathrm{sky},n} = \sqrt{\sigma_{\mathrm{sky},n}^2 + \frac{1}{N}(\sigma_{\mathrm{sky},n}^2 + \sigma_\mathrm{r}^2)}\;,
\label{eq:inflated-sky}
\end{equation}
where $N$ is the number of template observations combined 
into one subtraction image and $\sigma_{\mathrm{r}}$ is the readout noise per pixel.

\begin{table}
\caption{Observing characteristics assumed in our simulations: the 
telescope collecting area, the effective field of view, the CCD pixel size, its readout noise, 
the point spread function (PSF) radius and the rms uncertainty in the 
calibration.} 
\label{tab:survey-facts}
\centering
\begin{tabular}{|l|l|l|}
\hline
Collecting area                    & 39,507 $\mathrm{cm^2}$ &\tabularnewline
Effective FoV                      & 5.4 $\mathrm{deg^2}$ &\tabularnewline      
Pixel Size                         & 0.228 arcsec & (10$\mathrm{\mu m}$)\tabularnewline
Readout noise                      & 6 $\mathrm{e^-/pixel}$&\tabularnewline
PSF $\sigma$                       & 1.75 pixels & (0.4 arcsec)\tabularnewline
Calib. rms error                   & 0.04 mag &\tabularnewline                   
\hline
\end{tabular}
\end{table}

\subsection{Data quality cuts}
\label{sec:data-quality-cuts}

SN surveys usually impose various cuts on their data in order to ensure 
quality, and these cuts can be quite complex. \citet{Kessler09bmn}, for 
instance, required from the SDSS SNe at least: one measurement with 
signal-to-noise ratio (SNR) greater than 5 for each \emph{gri} filter; 
five measurements at SN rest-frame epoch $t$ (measured in days from 
maximum luminosity) in the range $-15<t<60$; 
one measurement at $t<0$; one measurement at $t>10$; and a $\chi^2$ fitting 
probability for the MLCS2k2 light curve model \citep{Jha07mn} greater than 
0.001. Unfortunately, given the big difference between the SDSS Supernova Survey 
and a narrow band survey, cuts cannot be transferred from one to the other, and 
therefore we must choose new cuts for selecting our simulated data. 

A simple and effective quality cut is to require a minimum number of measurements  
with SNR greater than some threshold, regardless of the filter or the epoch. 
Since there is a high number of such observations (up to 112) and they are 
scattered along the epochs and wavelengths, this cut automatically requires 
that the SN is observed in many filters and in different epochs. Besides, more 
complex and optimised quality cuts are likely to be dependent on a specific survey 
strategy, which is not the goal of this paper. Thus, we classified the SNe according 
to the number of measurements with $\mathrm{SNR}>3$ that they possess and put them 
into samples called `Group 20' (with a minimum of 20 such observations), `Group 30' 
(with a minimum of 30 such observations) and so on. The number of measurements with 
$\mathrm{SNR}>3$ is highly correlated with the number of measurements with 
$\mathrm{SNR}>5$ but provides a smoother selection. The quality group 30 provides 
a good balance between sample size and data quality and will receive most of 
our attention.

\subsection{SN typing and fitting process}
\label{sec:typing-fitting-process}

The SN typing was performed with the {\sc psnid} software \citep{Sako11mn} 
provided in the {\sc snana} package. This software basically compares the 
SN photometric measurements to a grid of templates which includes variations of 
SN type (Ia, Ibc and II), sub-types and parameters. 
A $\chi^2$ is computed for each point in this grid and is used to calculate the 
Bayesian probability that an observed SN belongs to one of the three types -- Ia, Ibc or II -- 
by marginalising over their sub-variations in the grid. In the case of a SN Ia, 
these sub-variations correspond to variations in the SALT2 parameters. For CC-SNe, 
they correspond to variations in the redshift $z$, the distance modulus $\mu$, the 
host extinction at \emph{V} band $A_V$, the ratio of total to selective extinction $R_V$, 
and to variations between different templates within that particular type. 
{\sc psnid} uses four type Ibc and four type II SN templates created from \citet{Nugent02mn} 
spectral templates, warped to match the photometry of eight spectroscopically 
typed nearby CC-SNe observed by SDSS. During the Bayesian 
probability calculations, {\sc psnid} uses the host galaxy redshift and its associated 
uncertainty as the mean and the standard deviation of a Gaussian prior for $z$. The priors 
for the other parameters were assumed flat. More details about the {\sc psnid} software 
can be found in \citet{Sako11mn}.

Our grid was built according to the ranges and intervals presented in 
Table \ref{tab:psnid-grid}. To classify a SN as type Ia, we required that its probability 
of belonging to this type should be above 0.9 (the sum of the three type probabilities is normalised 
to 1). Moreover, we required the $\chi^2$ $p$-value -- calculated for the best-fit SALT2 model -- to be greater than 0.01, 
so even if the type Ia model is the best fit for a light curve, 
it can still be ruled out as a bad fit. Given that {\sc psnid} CC-SN templates are not fully 
representative of all our simulated CC-SN light curves, to classify a SN as core-collapse 
we only required a 0.10 probability of it belonging to any core-collapse template.

\begin{table}
\caption{Ranges and number of nodes for the grid used by {\sc psnid} for typing SNe. Its 
parameters are the redshift $z$, the SN phase $t$ with respect to the time of maximum, 
a flux normalisation shift in mag $\Delta\mu$ from the value expected from fiducial cosmology, 
the SALT2 parameters $x_1$ and $c$, and the CC-SN host extinction parameters $A_V$ and $R_V$. 
All grid nodes are equally spaced with the exception of $z$ which is equally spaced in log scale.} 
\label{tab:psnid-grid}
\centering
\begin{tabular}{|c|c|c|c|}
\hline
Parameter   & Min. & Max. &  \# nodes \tabularnewline\hline
$z$         & 0.01 & 0.70 &  160     \tabularnewline
$t$         & -20  & 80   &  56      \tabularnewline
$\Delta\mu$ & -2.0 & 2.0  &  41      \tabularnewline
$x_1$       & -5.0 & 5.0  &  20      \tabularnewline
$c$         & -0.4 & 0.4  &  6       \tabularnewline
$A_V$       & -1.0 & 1.0  &  4       \tabularnewline
$R_V$       & 2.2  & 3.2  &  2       \tabularnewline
\hline
\end{tabular}
\end{table}

The fitting is performed by {\sc snana} through a $\chi^2$ minimisation using the 
{\sc minuit}\footnote{http://wwwasdoc.web.cern.ch/wwwasdoc/minuit/minmain.html} software.  
All our SALT2 model fits were performed with four free parameters since the SN redshifts 
were fixed to their host's photo-$z$s. In section \ref{sec:sne-photoz}, in particular, 
we perform a five free parameters fitting, leaving the SN redshift unconstrained.  

The estimate of the distance modulus is performed by solving Eq. \ref{eq:mB-def} for
$\mu$. However, the so-called ``nuisance parameters'' $\alpha$, $\beta$ and $M$ are 
usually not fixed by local measurements and are determined from the same data by 
minimising the scatter around an average distance for a particular redshift. In our 
analysis, this process was performed using the {\sc salt2mu} software \citep{Marriner11mn}, 
which assumes a fiducial cosmology and uses different average absolute magnitudes $M_i$ for 
each redshift bin $i$ to account for possible discrepancies. The parameters $\alpha$ and $\beta$ 
are determined by minimising the scatter in these bins, while $M$ is defined as the weighted 
average of $M_i$.   

\subsection{Broad band survey simulations}
\label{sec:broad-band-sim}

The SDSS-II SN survey simulation was performed 
using the standard SDSS characteristics as implemented in the {\sc snana} 
package and the same SN light curve models used for our narrow band 
simulations. Basically, the SDSS strategy consists of imaging the Stripe 82 
region (300 $\mathrm{deg^2}$) in the \emph{ugriz} filters every $\sim$ 4 
days, on average. We also applied the default {\sc snana} cuts required from the SDSS data:
\begin{enumerate}
\item at least three \emph{ugriz} filters with one or more observations with $\mathrm{SNR}>5$, in any epoch;
\item at least one observation made before the SN luminosity peak;
\item at least one observation made after ten days from the SN luminosity peak;
\item at least five observations made in different epochs.
\end{enumerate}

Two separate samples of simulated SDSS light curves were created, one 
associated with a redshift precision of 0.0005 (representing observations 
backed up by spectroscopy of the host galaxies) and another with a redshift 
precision of 0.03 (representing a pure broad band photometric survey).
The first case was termed spec-$z$ SDSS 
and the second one photo-$z$ SDSS. As in our fiducial survey, all 
redshift errors were assumed to be Gaussian. We did not consider any selection 
effects or sample size reductions that might be caused by spectroscopic 
follow-up of host galaxies, and the sole difference between these two 
simulated broadband data is the redshifts assigned to the SNe. In practise, 
however, a spec-$z$ broadband survey is likely to have its sample sizes 
reduced due to the scarcity of spectroscopic time. 

For comparing our narrow band survey outcomes with the DES SN survey we did not 
simulate DES light curves ourselves but used instead the results from 
\citet{Bernstein12mn}. 

\section{Results}
\label{sec:results}

The large area covered by our fiducial survey allows for a large number of SNe to be observed. 
Table \ref{tab:sample-sizes} presents the number of SNe that could be 
added to a catalogue every two months of searching, for our various quality groups. 
As a reference we also present the values obtained for the SDSS-II SN 
Survey simulation.  

\begin{table}
\caption{Number of SNe Ia and CC-SNe expected within each quality 
group, for every two months of search observations, assuming that the template 
observations have been made one month before them.\protect\footnotemark\; We only included 
SNe that passed light curve quality cuts and that were correctly typed. 
Results for an SDSS simulation were also included as a reference point.} 
\label{tab:sample-sizes}
\centering
\begin{tabular}{|l|c|c|c|c|c|}
\hline
Group     & SDSS & 20  & 30  & 50  & 70 \tabularnewline\hline 
\# SNe Ia & 330  & 760 & 500 & 210 & 75 \tabularnewline
\# CC-SNe &  60  & 120 &  90 & 50  & 25 \tabularnewline
\hline
\end{tabular}
\end{table}
\footnotetext{If the first two observations in each filter are 
performed one year or more before the last two and follow the same observation schedule, then both sets of 
observations can count as search observations.}

\begin{figure}
\includegraphics[width=1\columnwidth]{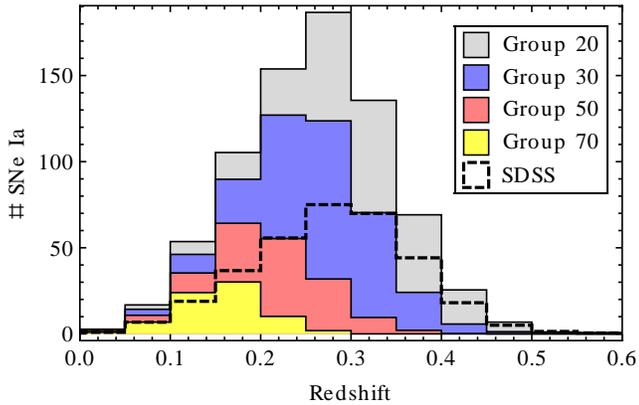}
\caption{SN Ia redshift distributions for our fiducial survey 
quality groups 20 (gray), 30 (blue), 50 (red) and 70 (yellow filling), 
superimposed in this order, from back to front. The number of SNe Ia was calculated for 
a season of two months of search imaging over 800 $\mathrm{deg^2}$. An increase in the 
minimum number of measurements with $\mathrm{SNR}>3$ decreases the average redshift 
and the number of SNe Ia that passes the cuts. Results obtained for the SDSS 
simulation (300 $\mathrm{deg^2}$), normalised to two months of search, 
are presented in thick, dashed contours (no filling).}
\label{fig:snia-z-hist} 
\end{figure}

Fig. \ref{fig:snia-z-hist} shows the expected redshift distribution of 
correctly typed SNe Ia for our fiducial survey under various selection cuts and 
for the SDSS simulation as a reference. 
The distributions for quality groups 20 and 30 populate a slightly shallower interval than 
the one obtained for SDSS, whereas the total number of SNe Ia 
is much bigger due to the larger area covered. As shown by \citet{Bernstein12mn}, 
DES will use a 4-m telescope and longer exposure times to identify up to 4000 SNe 
Ia with a redshift distribution peaking at $z\sim0.4$ and reaching $z\sim1.2$. 
The DES SN survey will use about 1300 h of observing time (approximately 0.32 of 
the total survey time), making an average of 1500 SNe Ia every two months of 
dedicated SN survey time and 470 SNe Ia every two months of total survey time.   

Fig. \ref{fig:snia-z-hist} also shows that an increase in the minimum number 
of observations with $\mathrm{SNR}>3$ required from the data reduces the sample 
sizes and redshift ranges. However, as presented in the following subsections, these 
reductions are accompanied by an increase in sample purity, photometry SNR and 
light-curve parameter recovery precision. An optimal balance can then be chosen 
according to the desired scientific goals. 

\subsection{Individual flux measurements}
\label{sec:flux-measurements}

The change from broad to narrow band filters modifies the qualitative behaviour of 
the survey, for instance by changing the error budget. As presented in Fig. 
\ref{fig:sky-read-noise}, the background noise from the sky is significantly 
reduced when compared to broad band, making the readout noise (often neglected 
in broad band imaging surveys) a relevant aspect of the survey. Moreover, since the 
calibration rms is a multiplicative flux error, it only contributes to the error budget 
at very low redshifts, while it might extend to higher redshifts for broadband 
surveys with the same exposure time. 

\begin{figure}
\includegraphics[width=1\columnwidth]{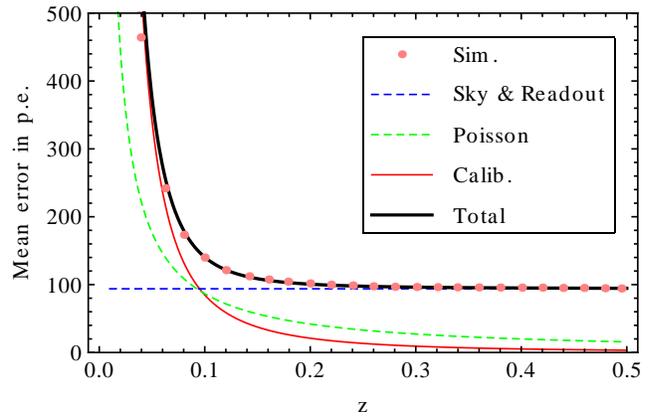}
\caption{Evolution of different error sources with redshift $z$, quantified by 
the CCD photon counting error (in photo-electrons, p.e.). The pink points 
represent the mean of the errors estimated by {\sc snana}, and the curves are 
simplified analytic models representing: the quadratic sum of SN and host galaxy Poisson noise (green, dashed 
line, decreasing with $z$), the quadratic sum of the readout and sky noise in 
the PSF (blue, dashed line, constant in $z$); the calibration rms contribution (solid, 
thin red line) and the total error (solid, thick black line). The simplified 
model represents the simulated data well, the Poisson noise is sub-dominant and 
the calibration error dominates the noise up to $z \sim 0.1$}.
\label{fig:error-vs-z} 
\end{figure}

In Fig. \ref{fig:error-vs-z} we compare the contributions from different error sources 
to the final flux measurement errors. The lines show simplified analytic error models 
which assume mean values for the zero point and for the sky noise and a fixed observer-frame 
absolute magnitude of -18.2 in all filters and redshifts. The points are the average of the 
results obtained from the detailed {\sc snana} simulation. We can notice that the 
Poisson noise from the SN and the host galaxy is almost always sub-dominant; 
and that the calibration rms contribution dominates up to redshifts $z \sim 0.1$. 

\begin{figure}
\includegraphics[width=1\columnwidth]{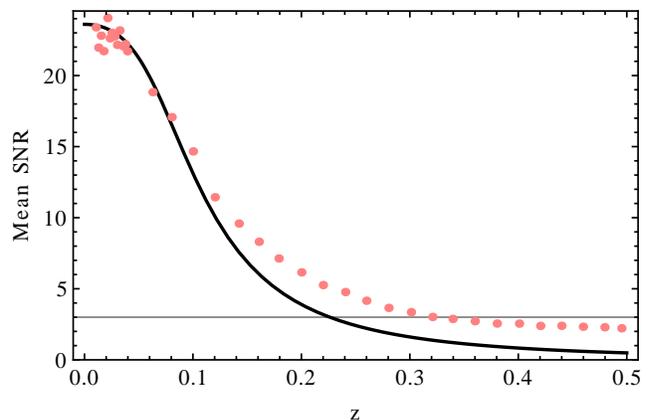}
\caption{Average SNR evolution with redshift $z$. The pink points 
represent the mean over simulated measurements with $\mathrm{SNR}>1$ for SNe Ia in group 30 (at least 30 
observations with $\mathrm{SNR}>3$). Our photometry toy model is represented by the 
thick black line. The horizontal gray line indicates the SNR=3 level. The SNR 
saturates at $\sim 24$ for low-$z$ due to the multiplicative error $\sigma_{ZP}$, 
and at $\sim 3$ for $z>0.3$ due to our selection cuts.} 
\label{fig:snr-vs-z} 
\end{figure}

The small bandwidth of the filters also affects the SNR by 
lowering the signal (see Fig. \ref{fig:snr-vs-z}). As expected, a narrow band survey 
will be shallower, maintaining a high SNR at lower redshifts. Selection effects are 
also expected to kick in a little before $z\sim 0.3$, when the average SNR reaches 
the level required (3 in this case) from some SN measurements. Due to the assumed 
calibration rms error of 0.04 mag, the SNR saturates, for low-$z$, at $\sim 25$. 

Fig. \ref{fig:snr-vs-z} also shows the output from our simplified analytic model 
which is detailed in Appendix \ref{sec:snr-model} (thick black line). It describes the 
general behaviour of the SNR reasonably well, although it underestimates the signal at higher redshifts. 
This is mainly due to effects that were ignored in our toy model: the drift with $z$ of the SN Ia 
luminosity peak from 4000 \AA\, to higher wavelengths (where the filter transmission is 
higher), an effect that can be accounted for with a K-correction; the time dilation 
of the light curves, that sustain detectable signals for longer periods; 
and selection effects such as the Malmquist bias. 
 
The results presented in Figs. \ref{fig:error-vs-z} and \ref{fig:snr-vs-z} are an 
average for all filters, and the specific results vary within the filter set. In 
general, there are three aspects that alter a narrow band filter's performance: 
its average transmission, the sky noise at its wavelength, and the SN Ia spectral 
energy density probed by the filter. 

An increment in the average transmission increases the signal, thus basically stretching 
the SNR curve in Fig. \ref{fig:snr-vs-z} along the horizontal axis (the sky noise is 
increased a bit as well). This effect benefits the intermediate wavelength filters 
(4500--8000 \AA, see Fig. \ref{fig:filter-transmission}). The sky noise will affect more 
strongly the reddest filters and those imaging the sky emission lines (see Fig. 
\ref{fig:sky-spectrum}). Finally, filters probing dimmer parts of the SN spectrum will 
also present a stronger drop in SNR with redshift. Since our SN Ia model around the epoch 
of maximum luminosity is brighter at (rest-frame) $\sim 4000$ \AA\, and quickly drops for lower wavelengths, 
this will mainly affect the bluest filters, specially since the spectrum will be 
stretched to higher wavelengths for higher redshifts. The final result for the average 
SNR per filter is presented in Fig. \ref{fig:snr-vs-flt}. The toy model in Appendix 
\ref{sec:snr-model} can be used to identify the effects of various survey 
characteristics on flux measurement errors and SNRs.

\begin{figure}
\includegraphics[width=1\columnwidth]{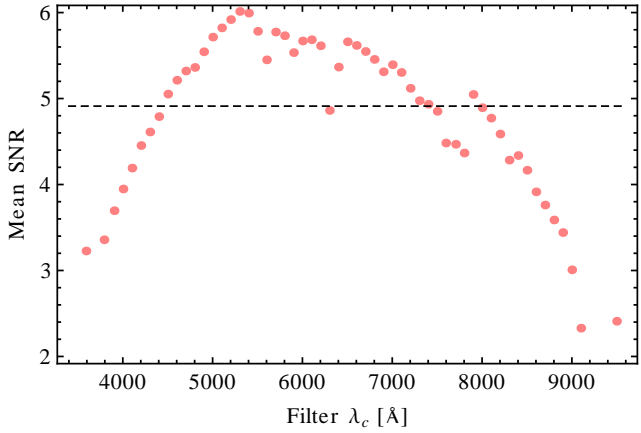}
\caption{Average SNR per filter. The pink points 
present the mean over simulated measurements with $\mathrm{SNR}>1$ for SNe Ia in group 30 (at least 30 
observations with $\mathrm{SNR}>3$).The dashed line indicates the average SNR for 
all filters. The effects of filter transmission functions, sky emission and 
different exposure times can be noted.} 
\label{fig:snr-vs-flt} 
\end{figure}

\begin{figure}
\includegraphics[width=1\columnwidth]{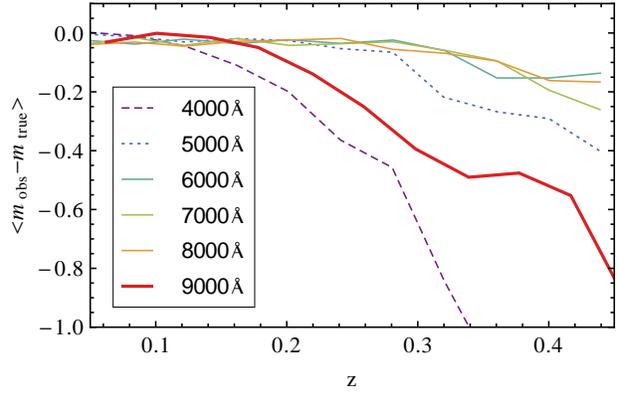}
\caption{Average difference between measured and true magnitudes for observations 
with $\mathrm{SNR}>1$ in filters numbers 4 (central wavelength $\lambda_{\mathrm{c}}=4000$ \AA, 
dashed purple line), 14 ($\lambda_{\mathrm{c}}=5000$ \AA, dotted blue line), 24 ($\lambda_{\mathrm{c}}=6000$ 
\AA, thin cyan line), 34 ($\lambda_{\mathrm{c}}=7000$ \AA, thin green line), 43 ($\lambda_{\mathrm{c}}=8000$ 
\AA, thin orange line) and 54 ($\lambda_{\mathrm{c}}=9000$ \AA, thick red line), as a function of 
redshift $z$. Filters on the outskirts of the wavelength range suffer stronger biases.} 
\label{fig:flux-bias} 
\end{figure}

Another relevant effect present in each individual flux measurement is a form of 
statistical Malmquist bias: as photon counting at the CCD is a statistical process, 
measurements near the selection threshold with positive fluctuations tend to be detected 
while those with negative fluctuations do not. This leads to an overestimation of the 
average photon emission from the source which should be taken into account if one 
is interested in measuring spectral features and flux ratios, for instance.
Fig. \ref{fig:flux-bias} shows this effect for six filters as a function of redshift, 
where we see that filters with lower average SNR (the very blue or very red) are the ones most affected. 
Apart from small fluctuations caused by the simulated sample finite size, the variations with redshift over the smooth 
dropping trend (better seen for the thick red curve) are caused by spectral features moving 
into and out of each filter's band. Curves for filters not shown in the plot can be roughly 
estimated by interpolating the plotted ones.

\begin{figure}
\includegraphics[width=1\columnwidth]{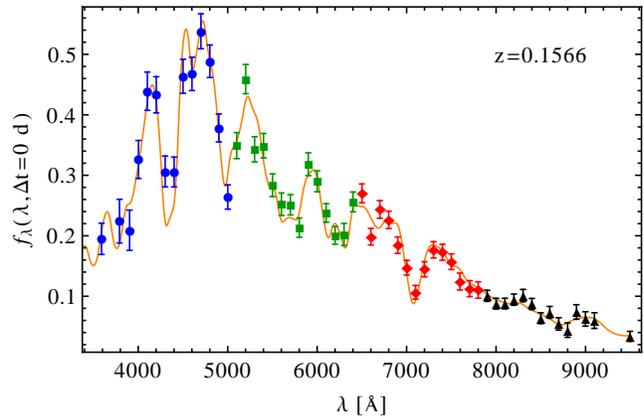}
\caption{Comparison between spectral template with Galactic extinction (orange line)  
and simulated measurements (data points) for a SN Ia at peak luminosity, at $z=0.157$, 
in the quality group 30. The average SNR for the plotted measurements is 12.1, 
and the spectrum units are arbitrary. Spectral features are clear. For brevity we show 
measurements on all filters, although in our fiducial strategy only a set of 14 different 
filters (presented as different markers and colours) would be observed on the same day.} 
\label{fig:low-z-spec} 
\end{figure}

\begin{figure}
\includegraphics[width=1\columnwidth]{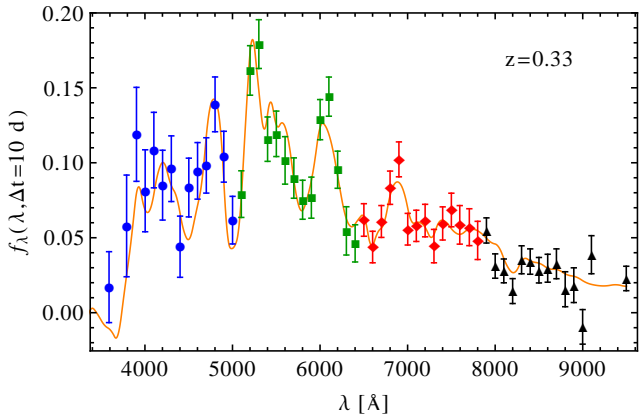}
\caption{Comparison between spectral template with Galactic extinction (orange line) 
and simulated measurements (data points) for a SN Ia 10 days after peak luminosity, at 
$z=0.33$, in the quality group 30. The average SNR for the plotted measurements 
is 4.9, and the spectrum units are arbitrary. Although noisy, large spectral 
features can still be detected. For brevity we show measurements on all 
filters, although in our fiducial strategy only a set of 14 filters (presented 
as different markers and colours) would be observed on the same day.} 
\label{fig:med-z-spec} 
\end{figure}

Figs. \ref{fig:low-z-spec} and \ref{fig:med-z-spec} compare observer frame 
SN Ia spectra near the epoch of maximum luminosity to simulated measurements. Both spectra and measurements include 
extinction by the Milky Way. To present a concise picture of the expected data quality we 
plotted the measurements on all 56 filters together in one epoch, but one should keep in mind 
that, in our fiducial strategy, the SN is observed in eight different epochs and in each one 
the observations are made on 14 contiguous filters (see Fig. \ref{fig:cadence-M01}). 

For low redshifts ($z<0.2$), the SN Ia spectral features are clear since 
they are much larger than the error bars (Fig. \ref{fig:low-z-spec}). In 
this redshift range, it is possible to detect 
and measure the light curves of nearly 170 SNe Ia every two months of search 
time (or every four months if the template imaging time is included). 
For higher redshifts ($z>0.3$), the measurements get noisier and 
this might prevent the detection of certain spectral features. However, 
as we show in section \ref{sec:parameter-recovery}, global light-curve 
parameters -- which are based on all 112 measurements -- can still be measured 
to high accuracy.

\subsection{SN typing}
\label{sec:typing-results}

We calculated the contamination fraction $\eta_{\mathrm{Ia}}$ of an SN Ia sample by 
CC-SNe using the formula:
\begin{equation}
\eta_{\mathrm{Ia}} = \frac{W_{\mathrm{CC}} N_{\mathrm{CC}}}{W_{\mathrm{Ia}} N_{\mathrm{Ia}} + W_{\mathrm{CC}} N_{\mathrm{CC}}},
\label{eq:contamination}
\end{equation}
where $W_X$ is the fraction of SNe of type $X$ that was identified as Ia and $N_X$ 
is the expected number of type $X$ SNe per month of search. Table \ref{tab:typing-results} 
shows that narrow band surveys can type SNe as well as broad band surveys, 
and that the performance is much higher for better quality groups. Estimates of SN Ia typing made by \citet{Bernstein12mn} 
indicate that DES will reach completeness $W_{\mathrm{Ia}}$ of $\sim 0.85$ and contamination 
$\eta_{\mathrm{Ia}}$ around 0.02.

\begin{table}
\caption{Typing performance for SDSS SNe with host galaxy photo-$z$ and spec-$z$, and for various 
quality groups of our fiducial survey. The columns present, from left to right: the 
sample, the average number of observable SNe Ia and CC-SNe per month of search, the fraction of SNe Ia 
and CC-SNe identified as Ia, and the final Ia sample contamination by CC-SNe.} 
\label{tab:typing-results}
\centering
\begin{tabular}{|l|c|c|c|c|c|}
\hline
Sample               & $N_{\mathrm{Ia}}$ & $N_{\mathrm{CC}}$ & $W_{\mathrm{Ia}}$ & $W_{\mathrm{CC}}$ & $\eta_{\mathrm{Ia}}$ \tabularnewline\hline
photo-$z$ SDSS       &  172           &    46          &   0.96         & 0.214          &  0.0562           \tabularnewline     
spec-$z$ SDSS        &  172           &    46          &   0.97         & 0.192          &  0.0503           \tabularnewline
Group 20             &  395           &   105          &   0.97         & 0.102          &  0.0274           \tabularnewline
Group 30             &  263           &    75          &   0.98         & 0.051          &  0.0148           \tabularnewline
Group 50             &  114           &    42          &   0.96         & 0.006          &  0.0023           \tabularnewline
Group 70             &   42           &    21          &   0.93         & $\la 0.002$    &  $\la 0.001$      \tabularnewline\hline
\end{tabular}
\end{table} 

Table \ref{tab:typing-results} also shows that the creation of an SN Ia sample is eased by the fact that 
its main sources of contamination -- the CC-SNe -- are dimmer than the SNe Ia (see Table 
\ref{tab:cc-sn-templates}) which reach absolute magnitudes of $M_B=-18.06$ or less 
\citep{Phillips93mn, Richardson02mn}. Thus, the CC-SNe populate lower redshifts, 
where the survey volume is smaller, and the number of detected CC-SNe is reduced. 
We can also see that photometric typing using broad bands can perform reasonably 
well, a result in agreement with other simulations \citep[e.g.][]{Campbell13mn} 
and with real data analysis \citep{Sako11mn}.

\subsection{SALT2 parameter recovery}
\label{sec:parameter-recovery}

Another way of analysing the quality of the SN data obtainable by a narrow band 
survey is to verify its precision on the recovery of the light-curve parameters 
used to simulate the data. However, it is important to keep in mind that a narrow 
band survey offers many more possibilities than can be simulated here. For instance, 
in the SALT2 model the colour variation is simply an extinction law without any 
implications to the SN spectra, thus it can be precisely measured with broad band 
filters and no new information is gained with a better wavelength resolution. 
The $x_1$ simulation and recovery with SALT2, on the other hand, is better suited for 
a narrow band survey analysis as it reflects variations both on light-curve width and 
spectral features. Still, it is possible that some spectral variations not present 
in the simulations could be detected by narrow band filters. Thus, this analysis is 
to be understood as a coarse, general guide to the survey's performance.

We selected all true SNe Ia that passed light curve cuts and were identified as 
Ias and binned them in redshift. For each SN we computed the difference 
$\Delta y=y_{\mathrm{f}}-y_{\mathrm{t}}$ between the fitted SALT2 parameter value $y_{\mathrm{f}}$ 
and the true one $y_{\mathrm{t}}$ ($y=m_B,\; x_1,\; c,\; t_0\;\mathrm{or}\;\mu$), and for each 
bin we computed the mean and the root mean square (rms) $\sigma_y$ of these differences. 
The mean can indicate the existence of any redshift dependent biases while the rms gives us 
a sense of the average error in each redshift bin. Their uncertainties were estimated as 
$\sigma_y/\sqrt{n}$ and $\sigma_y/\sqrt{2(n-1)}$, respectively, where $n$ is the number of 
SNe in that bin.

\begin{figure}
\includegraphics[width=1\columnwidth]{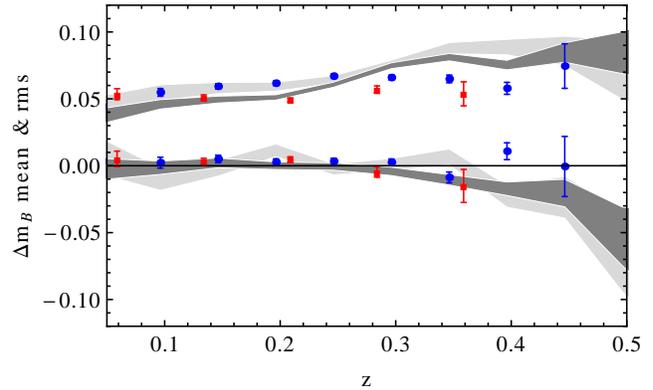}
\caption{Bias (bottom points) and rms (top points) of the difference between recovered and true $m_B$ 
SALT2 parameter for our narrow band survey groups 30 (blue circles) and 50 (red squares). The gray bands on the 
top part of the plot represent a $\pm2\sigma$ interval for the SDSS simulations rms and the bands on 
the bottom part represent $\pm2\sigma$ interval for the SDSS simulations biases. The light bands represent 
the photo-$z$ SDSS and the dark bands represent the spec-$z$ SDSS. All surveys have similar rms and 
biases.} 
\label{fig:bias-rms-mB} 
\end{figure}

Fig. \ref{fig:bias-rms-mB} shows the rms and bias calculated for the SN rest-frame apparent magnitude 
$m_B$ for our fiducial survey (quality groups 30 and 50) and for the SDSS simulations.\footnote{{\sc Snana} 
simulates the intrinsic scatter in the relation between distance and SN Ia observables (Eq. \ref{eq:mB-def}) 
by adding it to $m_B$ as extra scatter around its true value. 
Since we want here to assess the survey's precision in constraining the \emph{observed} 
$m_B$, we set $\sigma_{\mathrm{int}}=0$ for this particular analysis.} All 
surveys suffer from a bias which overestimates the SN luminosity at higher redshifts (the difference $\Delta m_B$ 
between recovered and true $m_B$ tends to be more negative), although it is less perceptible 
for the narrow band survey simulations. 
This is a form of statistical Malmquist bias, as explained in section \ref{sec:flux-measurements}.

\begin{figure}
\includegraphics[width=1\columnwidth]{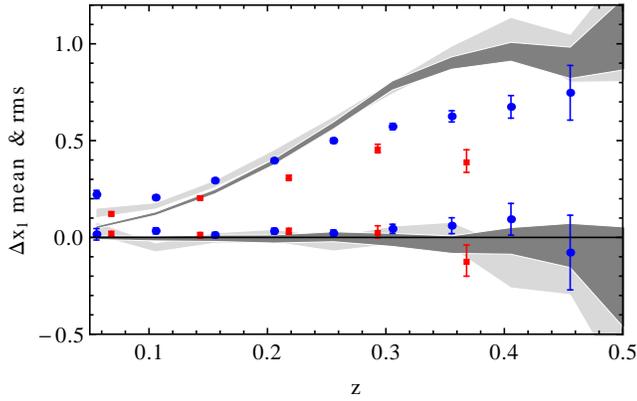}
\caption{Bias (bottom points) and rms (top points) of the difference between recovered and true $x_1$ 
SALT2 parameter for our narrow band survey groups 30 (blue circles) and 50 (red squares). As explained 
in Fig. \ref{fig:bias-rms-mB}, the gray bands represent the results for the SDSS simulations. The narrow 
band survey have smaller errors, as do higher quality groups.} 
\label{fig:bias-rms-x1} 
\end{figure}

On Fig. \ref{fig:bias-rms-x1} we notice that our fiducial survey can pin down more 
precisely the SALT2 $x_1$ parameter than our SDSS simulations, and that the increase 
in the data quality requirements also increases precision, as the $\Delta x_1$ rms 
is smaller for the quality group 50. The existence of a subtle constant bias favouring broader 
light curves (larger $x_1$) is possible, however this effect is very small -- maybe 
reaching $\sim 5$ per cent of the rms -- and is also insignificant for $\mu$ determination. 

\begin{figure}
\includegraphics[width=1\columnwidth]{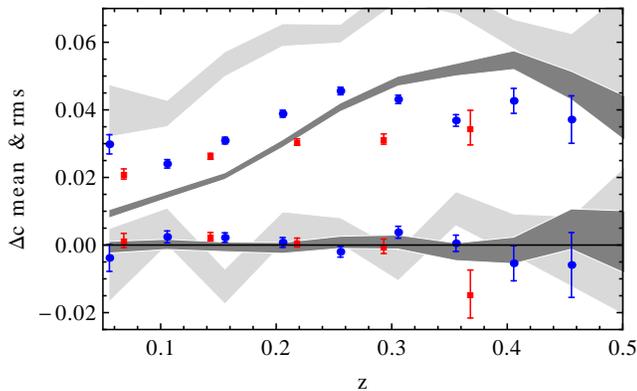}
\caption{Bias (bottom points) and rms (top points) of the difference between 
recovered and true $c$ SALT2 parameter for our narrow band survey groups 30 
(blue circles) and 50 (red squares). As explained in Fig. \ref{fig:bias-rms-mB}, 
the gray bands represent the results for the SDSS simulations. A strong 
redshift prior helps reducing the average errors, and broad band photometry is 
good at constraining $c$ as long as it is backed up with spectroscopic redshifts 
of the host galaxies.} 
\label{fig:bias-rms-c} 
\end{figure}

As suggested above, the advantages of narrow band filters are not as significant for 
constraining SALT2 colour $c$. Fig. \ref{fig:bias-rms-c} shows that spec-$z$ 
SDSS can perform as well as the narrow band quality group 30, on average, and better than both 
quality groups at low redshifts. DES, when backed up by spectroscopy, reach a 
colour rms of 0.031 in the range $0.2<z<0.4$ and an average of 0.046 for its 
full sample \citep{Bernstein12mn}. 
It is also possible to notice that $c$ measurements are severely affected by a looser 
redshift prior, as shown by the photo-$z$ SDSS much larger rms. This is expected since 
a change in redshift drifts the SN spectrum and changes the expected flux in each 
observer-frame filter. Therefore, even if differences in colour are simply broad 
band features, pure narrow band surveys still yield better colour measurements than pure 
broad band surveys given their much better photo-$z$ constraints.

\begin{figure}
\includegraphics[width=1\columnwidth]{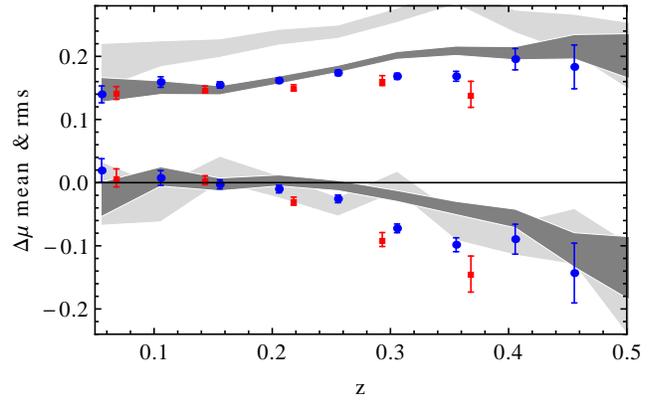}
\caption{Bias (bottom points) and rms (top points) of the difference between recovered and true 
distance modulus $\mu$ for our narrow band survey groups 30 (blue circles) and 50 (red squares). 
As explained in Fig. \ref{fig:bias-rms-mB}, the gray bands represent the results for the SDSS 
simulations. All simulations suffers from biases and, apart from the photo-$z$ SDSS, 
all simulations are very close to the 0.14 intrinsic scatter.} 
\label{fig:bias-rms-mu} 
\end{figure}

Finally, the quality of distance measurements with narrow band surveys is high, 
as its rms stays close to the SN Ia intrinsic scatter of 0.14 mag we assumed for our 
simulations (see Fig. \ref{fig:bias-rms-mu}). The DES simulations have $\sigma_{\mathrm{int}}=0.13$ 
and reached a $\Delta\mu$ rms of 0.16 in the range $0<z<0.5$ and an average of 0.20 
for the full DES sample. It is also possible to see how the large uncertainty in $c$ 
caused by the loose redshift prior in the photo-$z$ SDSS simulation affects the 
distance measurements.

Fig. \ref{fig:bias-rms-mu} also shows that all our simulations are affected by a bias that 
underestimates distances at large redshifts. This bias results from a combination of 
the $m_B$ bias presented in Fig. \ref{fig:bias-rms-mB} and from a classical Malmquist bias 
of its own. The distance modulus $\mu$ is calculated from Eq. \ref{eq:mB-def} by setting 
$m_B$, $x_1$ and $c$ to the measured values and $\alpha$, $\beta$ and $M$ to values that 
minimise the sample's scatter around the distance predicted by a particular cosmology. 
However, this relation between $\mu$ and $m_B$, $x_1$ and $c$ is not perfect and this 
imperfection is modelled by the intrinsic scatter. Given that for fixed values of $\mu$, 
$x_1$ and $c$ the SNe Ia still present intrinsic luminosity variations, observations near 
the threshold will preferentially detect brighter objects, thus giving the impression of 
a smaller distance. For cosmological studies, this bias has to be corrected by simulations. 
Table \ref{tab:avg-rms-M01} summarises the precision attainable in the SALT2 parameters 
by each SN Ia sample. We verified that the levels of CC-SNe contamination estimated 
in section \ref{sec:typing-results} are too small to affect the determination of the 
nuisance parameters $\alpha$, $\beta$ and $M$ and, therefore, the distance inferred from 
SNe Ia.

\begin{table}
\caption{The rms of the differences between fitted and true SALT2 parameters. No binning in 
redshift was performed.}
\label{tab:avg-rms-M01}
\centering
\begin{tabular}{|c|c|c|c|c|c|c|c|}
\hline
Group               & $\sigma_{m_B}$ & $\sigma_{x_1}$ & $\sigma_c$ & $\sigma_{t_0}$ & $\sigma_{\mu}$ \tabularnewline\hline
photo-$z$ SDSS      & 0.074          & 0.72          & 0.066     & 0.87          & 0.25          \tabularnewline
spec-$z$ SDSS       & 0.069          & 0.69          & 0.043     & 0.77          & 0.19          \tabularnewline
20                  & 0.074          & 0.61          & 0.054     & 1.00          & 0.18          \tabularnewline
30                  & 0.063          & 0.47          & 0.040     & 0.71          & 0.16          \tabularnewline
50                  & 0.052          & 0.30          & 0.029     & 0.48          & 0.15          \tabularnewline
70                  & 0.046          & 0.21          & 0.024     & 0.42          & 0.14          \tabularnewline
\hline
\end{tabular}
\end{table}

\subsection{SNe photo-$z$ fitting}
\label{sec:sne-photoz}

In our main analysis of SALT2 parameter recovery we fixed the SNe redshifts 
to their host galaxies photo-$z$s, and in the analysis of SN typing with {\sc psnid} 
we used the host galaxies photo-$z$s as Gaussian redshift priors; in 
both cases, the errors on the host galaxies photo-$z$s were Gaussian with 
$\sigma_z=0.005$. In this section we briefly investigate the data outcome for SNe without 
including any information from their hosts. This translates into typing the SNe using a 
flat redshift prior and into doing a five-parameter instead of a four-parameter SALT2 
fitting (the SN redshift is now a free parameter that can also be tested for recovery 
precision, and which we call ``SN photo-$z$'').

\begin{figure}
\includegraphics[width=1\columnwidth]{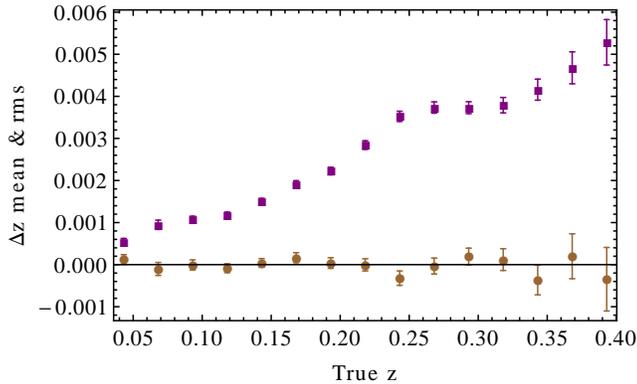}
\caption{Bias (bottom brown circles) and rms (top purple squares) of the difference 
between recovered and true redshifts for our narrow band survey quality group 30 when 
the SALT2 fitting is performed with the redshift as a free parameter and no host photo-$z$ 
information is used. In this plot we also removed outliers (SNe Ia with $\Delta z > 4\sigma_z$, 
corresponding to $\sim$ 2 per cent of the sample). Our simulations can achieve very low rms and 
no significant bias for the majority of SNe Ia.} 
\label{fig:sn-photo-z} 
\end{figure}

The SN photo-$z$ distribution obtained includes a small fraction of outliers 
(SNe with photo-$z$s more than 4$\sigma$ away from their true values), 
whose absolute difference between recovered and true redshifts commonly 
surpass 0.1. However, this fraction is very low, being 0.038 for the quality group 
20 and reaching 0.007 for quality group 50. On top of that, the remaining SNe Ia 
have extremely accurate photo-$z$s, presenting a symmetrical error 
distribution, rms below 0.005 and no noticeable bias (Fig. \ref{fig:sn-photo-z} 
shows the SN photo-$z$ bias and rms for the quality group 30).   
This precision makes sense as narrow band filters can clearly detect SN spectral features 
(see Fig. \ref{fig:low-z-spec} and \ref{fig:med-z-spec}). 
In comparison, \citet{Kessler10bmn} and \citet{Sako11mn} showed both 
with simulations and real data that SDSS SN photo-$z$ in the same 
redshift range is not free from bias and reach an average rms of 
$\sim 0.03$ or more.   
Notice that even though our simulations can reach very small photo-$z$ errors, 
in practise these are limited to 0.005 by intrinsic uncertainties such as the 
rms between SN and host galaxy redshifts \citep{Kessler09bmn}.

The use of the five-parameter SALT2 fit, however, introduces significant biases in all the 
other parameters: the SN Ia colour, for instance, was on average measured to be redder 
than in reality by 0.005 mag. A similar bias was already reported by \citet{Olmstead13mn} 
for an analysis of SDSS SN data and was attributed to a bias in the SN photo-$z$ and 
its degeneracy with colour. To test if the SN photo-$z$ values are responsible for these 
biases we fixed them as the SN redshifts and reran the fitting, this time with four free 
parameters. All biases then disappeared, indicating that the five parameter fitting method 
might be responsible for them. This issue still needs further investigation.

Lastly, Table \ref{tab:typing-no-host-z} shows that the lack of a Gaussian redshift prior 
made the narrow band typing performance slightly worse than before; however, it remained comparable to 
(or better than) those obtained for the SDSS simulations.

\begin{table}
\caption{Typing performance for our fiducial survey when a flat redshift prior is 
used instead of the Gaussian prior from the host photo-$z$. The columns are the 
same as in Table \ref{tab:typing-results}.} 
\label{tab:typing-no-host-z}
\centering
\begin{tabular}{|l|c|c|c|c|c|}
\hline
Sample               & $N_{\mathrm{Ia}}$ & $N_{\mathrm{CC}}$ & $W_{\mathrm{Ia}}$ & $W_{\mathrm{CC}}$ & $\eta_{\mathrm{Ia}}$ \tabularnewline\hline
photo-$z$ SDSS       &  172           &    46          &   0.96         & 0.214          &  0.0562           \tabularnewline     
spec-$z$ SDSS        &  172           &    46          &   0.97         & 0.192          &  0.0503           \tabularnewline
Group 20             &  395           &   105          &   0.95         & 0.1519          &  0.0406           \tabularnewline
Group 30             &  263           &    75          &   0.97         & 0.0984          &  0.0284           \tabularnewline
Group 50             &  114           &    42          &   0.96         & 0.0152          &  0.0057           \tabularnewline
Group 70             &   42           &    21          &   0.96         & 0.0020          &  0.0010           \tabularnewline
\hline
\end{tabular}
\end{table}

\section{Optimising the survey}
\label{sec:optimization}

In this section we look into possible ways of improving narrow band SN data, 
specially without requiring better instruments. It is clear that a larger light 
collecting area and lower noise levels will improve the data, even if in different 
ways. As exemplified by our photometry toy model presented in Appendix 
\ref{sec:snr-model}, the effect of a larger mirror, better filter transmission, 
more exposure time and larger bandwidths are all the same in terms of increasing 
photometry SNR and redshift depth. A larger exposure time, however, results 
in a loss of sky area covered during an observing season (presenting a trade-off 
between SNR and number of SNe), while larger bandwidths result in a loss in spectral resolution. 
Both of these changes might be beneficial depending on the survey's goals.

Although less noisy data are always better, noise reduction might result in bad trade-offs 
or might yield very little gain. As presented in Fig. \ref{fig:error-vs-z}, 
even though the calibration rms error dominates the error budget at low 
redshifts and limits the increase of the SNR, the number of SNe affected by it is small since 
the survey volume at low redshifts is small. Thus, to improve the SNR for a large number of 
SNe, we should pay attention to the signal and to the dominant noise sources at higher 
redshifts (sky and CCD readout). 
 
\begin{figure}
\includegraphics[width=1\columnwidth]{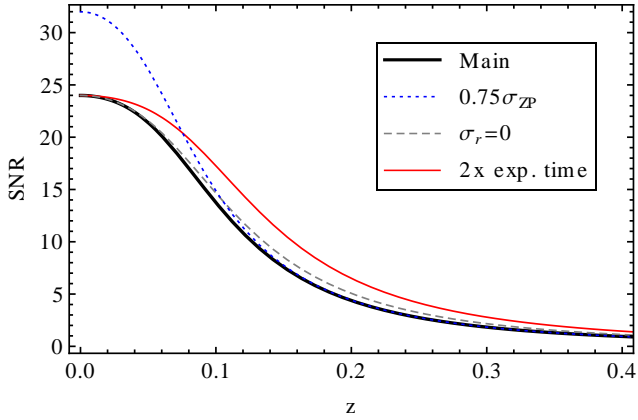}
\caption{Estimate of the average SNR at each redshift $z$ for our fiducial SN survey (black 
thick line), calculated with the toy model described in Appendix \ref{sec:snr-model}. We compare 
it with a survey with improved calibration rms (25 per cent decrease in $\sigma_{ZP}$, dotted blue line), 
no readout noise ($\sigma_{\mathrm{r}}=0$, dashed gray line), and two times longer exposure times 
(solid red line). A lower multiplicative error $\sigma_{ZP}$ greatly improves 
data (but only at low-$z$) while decreasing CCD readout noise does not. More exposure time 
improves data at all redshifts.} 
\label{fig:snr-compare} 
\end{figure}

Fig. \ref{fig:snr-compare} compares the expected effects of improving calibration precision 
(in terms of reducing the zero point rms), decreasing the CCD readout noise and doubling the 
exposure time. A better calibration is highly beneficial for SNe at $z\la 0.1$, and such 
improvement might be worthwhile if one is interested in these objects (even though, as 
Fig. \ref{fig:snia-z-hist} points out, the amount of SNe Ia at $z<0.1$ is small). It is 
also possible to notice that the yield from reducing the readout noise is very small, even 
with an impossible $\sigma_{\mathrm{r}}=0$, as the sky noise is already comparable to 
$\sigma_{\mathrm{r}}=6\mathrm{e^-/pixel}$ (see Fig. \ref{fig:sky-read-noise}) and would 
dominate the total noise at $z\ga 0.1$ if $\sigma_{\mathrm{r}}$ was reduced.

Fig. \ref{fig:snr-compare} also shows that increasing the exposure time would be 
beneficial to SNe at all redshifts. Keeping the total survey time constant, this 
increase can be achieved by reducing: (a) the area observed in one season; 
(b) the number of filters; (c) the cadence (number of times each field is observed 
in a given period of time); or (d) the so-called overhead time (time wasted, during 
observing hours, to read the CCDs and to reposition the telescope). Item (a) results 
in a simple trade-off with sample size and will not be further investigated here. 
Item (b) is analysed in section \ref{sec:no-red-filters}, while the effect of increasing 
the cadence [the reverse of item (c)] is studied in section \ref{sec:big-cadence}. We 
investigate the effect of item (d) in section \ref{sec:overhead-time}, and in the following 
subsection we present an optimisation method that does not involve increasing the SNR.

\subsection{Dispersed observations}
\label{sec:dispersed-obs}

\begin{figure}
\includegraphics[width=1\columnwidth]{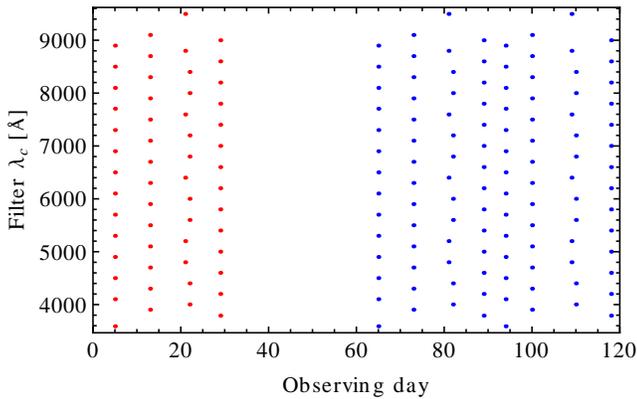}
\caption{Example of observation schedule for our scenario with dispersed observations. 
The red and blue points represent the template and search observations, respectively 
(the template observations do not necessarily need to be dispersed). Four different sets 
of 14 filters are observed twice during template and twice during search observations. 
In each set, the filters are equally spaced in wavelength.} 
\label{fig:cadence-M05} 
\end{figure}

An interesting approach to improve SN data quality is to redistribute the observations 
among the epochs or the spectrum while maintaining the same SNR level for the individual flux 
measurements. Fig. \ref{fig:cadence-M05} shows the observation schedule for this new scenario. 
In each epoch, the observations are evenly spread over the 56 filters set wavelength range. 
As in our main scenario (Fig. \ref{fig:cadence-M01}), only 14 filters are observed in each 
epoch, the search epochs are evenly distributed over $\sim 2$ months and each filter is observed in 
the 2+(1+1) strategy. 

\begin{figure}
\includegraphics[width=1\columnwidth]{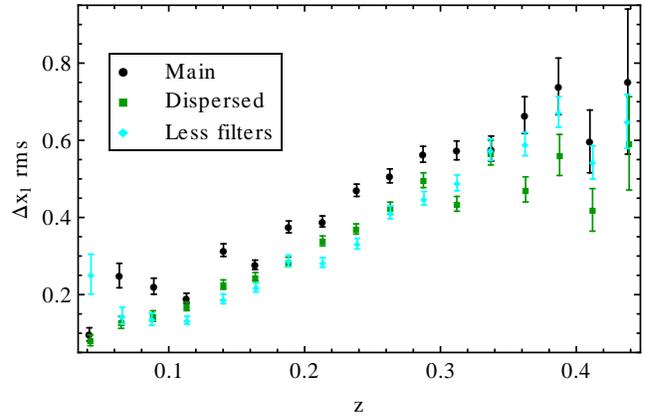}
\caption{Rms of the difference between recovered and true $x_1$ SALT2 parameter for 
quality groups 30 of the main scenario (black circles), the scenario with 
dispersed observations (green squares) and the scenario with less filters and more 
exposure time (cyan diamonds), as a function of redshift. Dispersed 
observations and more exposure time on fewer filters provide significantly better 
precision.} 
\label{fig:M02-M05-x1-rms} 
\end{figure}

By repeating the analysis with this new observation schedule, our simulations show 
that our ability to constrain SALT2 parameters is significantly enhanced, 
specially for lower quality groups -- which present more 
room for improvement. Colour is the most strongly improved parameter, probably due to 
the increased leverage of sampling the whole wavelength range in each epoch. Table 
\ref{tab:rms-M05} summarises the average SALT2 parameters uncertainties for 
this scenario and Fig. \ref{fig:M02-M05-x1-rms} compares, as an example, 
its $\Delta x_1$ rms redshift dependence for quality group 30 with the one obtained 
for the main scenario. The redshift distribution of the SNe Ia remained similar, 
as well as the bias on SALT2 parameters and on the distance modulus, although the 
subtle bias favouring larger $x_1$ got smaller in this scenario. 
 
\begin{table}
\caption{The rms of the differences between fitted and true SALT2 parameters for 
the scenario with dispersed observations. No binning in redshift was performed.}
\label{tab:rms-M05}
\centering
\begin{tabular}{|c|c|c|c|c|c|c|c|}
\hline
Group               & $\sigma_{m_B}$ & $\sigma_{x_1}$ & $\sigma_c$ & $\sigma_{t_0}$ & $\sigma_{\mu}$ \tabularnewline\hline
photo-$z$ SDSS      & 0.074          & 0.72          & 0.066     & 0.87          & 0.25          \tabularnewline
spec-$z$ SDSS       & 0.069          & 0.69          & 0.043     & 0.77          & 0.19          \tabularnewline
20                  & 0.061          & 0.50          & 0.039     & 0.88          & 0.17          \tabularnewline
30                  & 0.054          & 0.39          & 0.032     & 0.63          & 0.16          \tabularnewline
50                  & 0.046          & 0.27          & 0.023     & 0.44          & 0.15          \tabularnewline
70                  & 0.044          & 0.18          & 0.017     & 0.32          & 0.15          \tabularnewline
\hline
\end{tabular}
\end{table} 

\begin{figure}
\includegraphics[width=1\columnwidth]{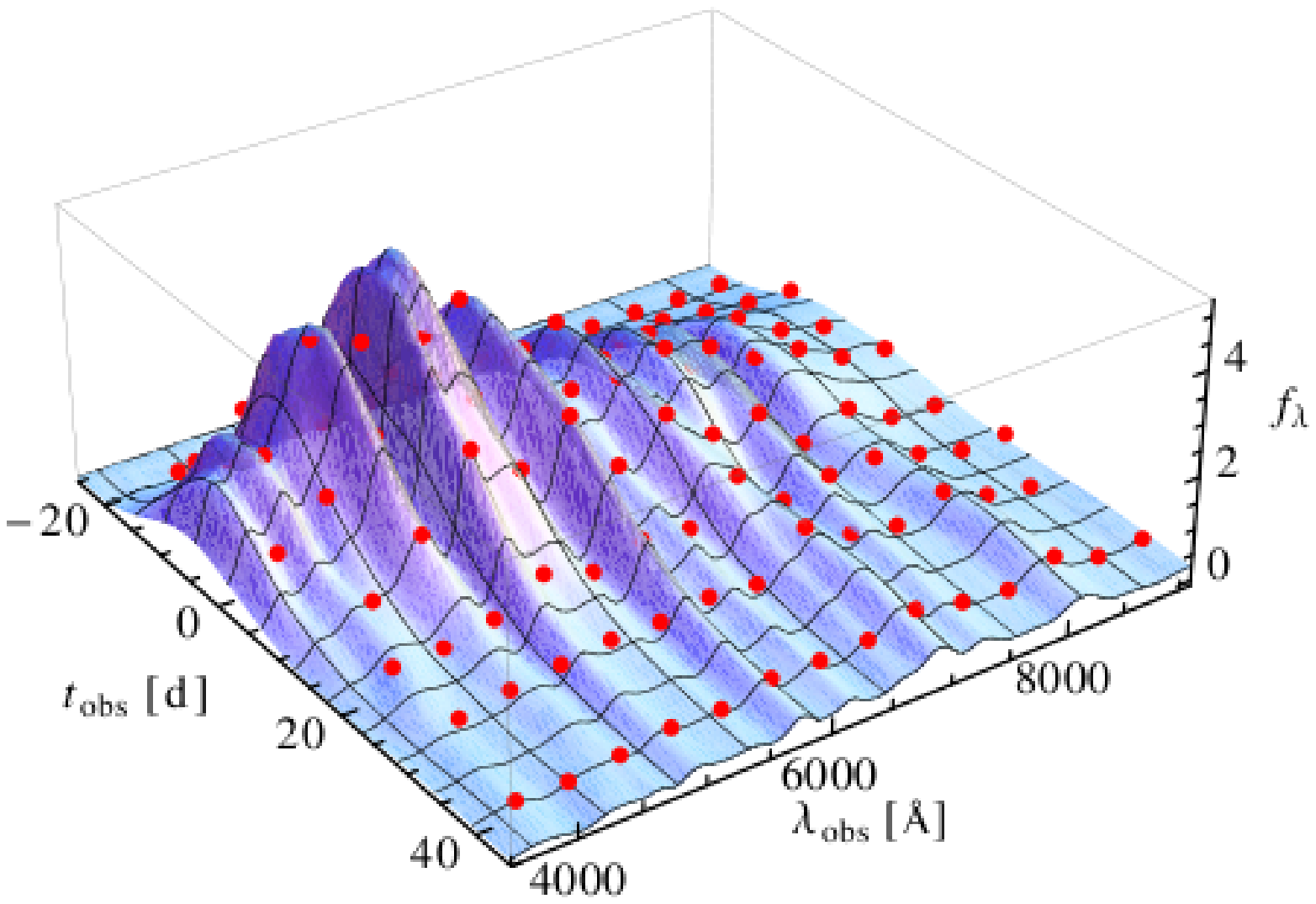}
\caption{SN Ia spectral surface $f_{\lambda}(\lambda_{\mathrm{obs}},t_{\mathrm{obs}})$ 
at $z=0.25$, convolved with a top-hat function 100 \AA\, wide, 
in arbitrary flux units. The time $t_{\mathrm{obs}}$ axis is given 
in days from maximum luminosity. The red points depict measurements in 
54 narrow band filters following the strategy with dispersed observations 
(errors not included). A good sampling of $f_{\lambda}(\lambda_{\mathrm{obs}},t_{\mathrm{obs}})$ 
can better constrain its parameters.} 
\label{fig:spectral-surface} 
\end{figure}

This improvement on constraining light-curve parameters is easy to understand if 
we remember that we are trying to constrain a spectral surface 
$f_{\lambda}(\lambda_{\mathrm{obs}},t_{\mathrm{obs}})$ (Eq. \ref{eq:salt2-flux}): if our 
measurements are better spread over this surface, we have a better idea of 
its shape (see Fig. \ref{fig:spectral-surface} for a helpful representation of this idea). 
It is true that some regions of the spectra might vary more and 
thus contain more information, but these region's location change with redshift. 
Therefore, an even sampling of $f_{\lambda}(\lambda_{\mathrm{obs}},t_{\mathrm{obs}})$ might 
be the best option for constraining its parameters. It is important to 
remember that although this strategy is better for describing overall 
characteristics of the light curves, one looses information about specific 
spectral features that might be measured within our main scenario.

\subsection{Less filters, more time}
\label{sec:no-red-filters}

By analysing SN spectra, one notices that the most luminous parts and 
many important features for typing SNe (H, He and SiII lines) lie below 
$\sim 6400$ \AA\, and only enter the reddest filters ($\lambda\ga 8000$\AA) 
at redshifts $z\ga 0.25$, when our SN Ia redshift distribution starts declining. 
Therefore, for our survey's depth, these filters convey little information 
about supernovae, and their allocated time might be put to better use if 
distributed among the other filters.

We created a new scenario in which the reddest 14 filters (which, in our main 
scenario, had twice the regular exposure time -- see section 
\ref{sec:fiducial-survey}) were removed and their time evenly distributed 
to the other filters. This filter removal also saved overhead time, and 
we were able to increase the remaining filter's exposure time by 73 per cent. 

When comparing this scenario with our main scenario, it is important to 
keep in mind that the same requirements in terms of number of observations with 
$\mathrm{SNR>3}$ result in a more restrictive selection for the scenario with 
less filters since the chance of achieving a certain number of good observations 
is smaller when the total number of observations is smaller. Thus, the best 
way of comparing the results is to remember the trade-off between number of SNe and 
data quality and take both into account. 

\begin{figure}
\includegraphics[width=1\columnwidth]{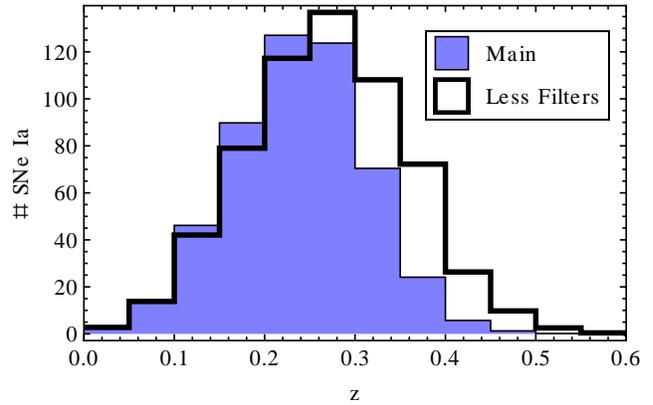}
\caption{SN Ia redshift distribution for the main scenario (thin 
contours, blue filling) and for the scenario with less filters (thick contours, 
no filling). Both histograms are for two months of search observations and quality 
group 30. The decrease in the number of filters and the corresponding extension of 
exposure time increased the survey's depth and the total number of SNe Ia from 
500 to 610 every two months of search.} 
\label{fig:zdist-M02-M06} 
\end{figure}

As expected, the increase in exposure time made the survey more deep and massive 
(see Fig. \ref{fig:zdist-M02-M06}). Table \ref{tab:rms-M06} also shows that the precision in the recovery 
of SALT2 parameters improved for $x_1$, $c$ and $m_B$, whereas it got slightly 
worse for $t_0$ -- probably due to the smaller number of observing epochs -- and 
remained practically the same for $\mu$ since it is limited by the intrinsic 
scatter. As an example, Fig. \ref{fig:M02-M05-x1-rms} shows the redshift 
dependence of the $\Delta x_1$ rms for this scenario. The typing 
performance remained basically the same.

\begin{table}
\caption{The rms of the differences between fitted and true SALT2 parameters for 
the scenario without the 14 reddest filters. No binning in redshift was performed.}
\label{tab:rms-M06}
\centering
\begin{tabular}{|c|c|c|c|c|c|c|c|}
\hline
Group               & $\sigma_{m_B}$ & $\sigma_{x_1}$ & $\sigma_c$ & $\sigma_{t_0}$ & $\sigma_{\mu}$ \tabularnewline\hline
photo-$z$ SDSS      & 0.074          & 0.72          & 0.066     & 0.87          & 0.25          \tabularnewline
spec-$z$ SDSS       & 0.069          & 0.69          & 0.043     & 0.77          & 0.19          \tabularnewline
20                  & 0.064          & 0.58          & 0.056     & 1.00          & 0.20          \tabularnewline
30                  & 0.057          & 0.43          & 0.037     & 0.78          & 0.17          \tabularnewline
50                  & 0.051          & 0.27          & 0.024     & 0.53          & 0.16          \tabularnewline
70                  & 0.046          & 0.16          & 0.017     & 0.39          & 0.14          \tabularnewline
\hline
\end{tabular}
\end{table}

\subsection{Less SNR, more cadence}
\label{sec:big-cadence}

Due to the transient nature of SNe, a higher spectral surface sampling rate in 
time yields better constraints to its shape. For a fixed instrument, this increase in 
cadence is achieved by saving observing time either by reducing the area imaged 
or by reducing the time spent in each individual exposure. Whereas the first 
option clearly results in a trade-off between number of SNe observed and 
light curve measurement quality, in principle it is not obvious what the effect 
of the second option would be: while each individual measurement would have a smaller 
SNR, the amount of independent measurements would be higher.

To test this last option, we simulated SN observations where each filter 
was imaged four times instead of two, while the individual exposure times 
were reduced from 60 to 23.9 s for filters number 1--42 and from 120 to 53.9 s 
for filters number 43--56. These exposure times were chosen so as to keep constant the 
observed area of the sky (note that the increase in the number of exposures also 
increases the amount of wasted overhead time).

In terms of typing efficiency and recovery of SALT2 parameters, this simulation 
presented basically the same performance as our fiducial strategy, indicating that 
a larger number of observations can compensate for a smaller individual SNR, at least 
in the range tested. However, the SNR reduction decreases the survey depth, thus making 
the overall performance worse for this scenario.
 
\subsection{Overhead time reduction}
\label{sec:overhead-time}

Assuming that the overhead time $t_{\mathrm{o}}$ is dominated by CCD readout time $t_{\mathrm{r}}$, 
one can trade low $\sigma_{\mathrm{r}}$ and high $t_{\mathrm{o}}$ for high $\sigma_{\mathrm{r}}$ and 
low $t_{\mathrm{o}}$ since $\sigma_{\mathrm{r}}$ and $t_{\mathrm{r}}$ follow a power-law relation:  

\begin{equation}
\sigma_{\mathrm{r}}=\left( 2+\frac{50\mathrm{s}}{t_{\mathrm{r}}}\right) \mathrm{e^-/pixel}. 
\label{eq:r-noise-time}
\end{equation}
The relation above was based on \citet{Jorden12mn} and adjusted to match 
$\sigma_{\mathrm{r}}(t_{\mathrm{r}}=12\mathrm{s})=6\mathrm{e^-/pixel}$. The time 
saved from reading the CCD could then be used to increase the exposure time 
and therefore the flux signal. This potential option for improving the SNR 
was investigated only through the use of our photometry toy model. 

\begin{figure}
\includegraphics[width=1\columnwidth]{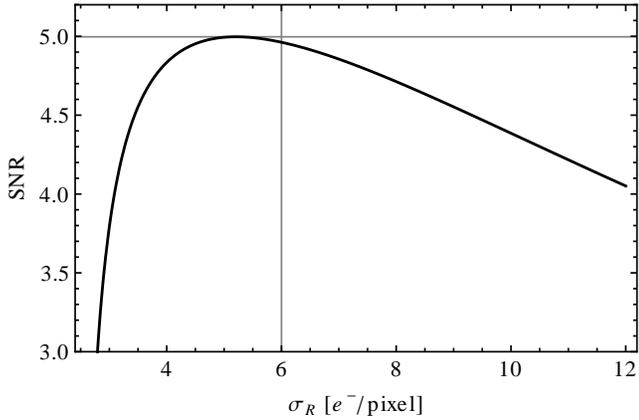}
\caption{Average SNR for a SN Ia at $z=0.19$, calculated for a narrow band survey 
using the toy model from Appendix \ref{sec:snr-model}, as a function of the CCD 
readout noise, assuming that observing hours are fixed and split between exposure time 
and CCD readout time and that readout noise and time follow the relation presented by 
Eq. \ref{eq:r-noise-time}. Vertical and horizontal gray lines are shown as references. 
The relations for different redshifts are very similar.} 
\label{fig:snr-rnoise} 
\end{figure}

Fig. \ref{fig:snr-rnoise} shows how the average SNR responds to such trade, 
assuming that the time saved from CCD reading is used to increase exposure time. 
Although it is beneficial to increase $\sigma_{\mathrm{r}}$ in some regimes, 
our fiducial value is close to the optimum and not much can be gained from 
this trade.

\section{Systematic uncertainties}
\label{sec:systematics}

Our choice of systematic uncertainty sources to be studied was based on 
the list presented by \citet{Bernstein12mn} for the DES supernova 
simulations: (a) offsets on the filter zero points; (b) offsets on the filter central 
wavelengths; (c) contamination by CC-SNe; (d) an error on the priors adopted for 
dust extinction; and (e) bias on inter-calibration with low redshift SN Ia samples. 
However, many of these sources are not intrinsic to the instrument and filters used: 
item (d) only applies to the MLCS2k2 model and item (e) involves the combination with 
other data sets. The contamination by CC-SNe (c) might depend on the instrument and 
strategy as it depends on selection effects, however it impacts specific uses of SN Ia 
samples -- like measuring the equation of state of dark energy -- and not the individual 
measurements or the recovery of SALT2 parameters. Moreover, \citet{Bernstein12mn} showed 
that the systematic uncertainty caused by a contamination level similar to ours was 
sub-dominant, so we focused our analysis 
on the effects of offsets (a) on the filter zero points and (b) on the filter central 
wavelengths. We also analysed the effects of biases in the photo-$z$ in section 
\ref{sec:host-z-bias} as they may be relevant for our particular survey.

\subsection{Filter central wavelengths}
\label{sec:central-wavelength}

In practise, the filter set used to image the SNe will not be exactly like the synthetic
transmission curves we use to compute the expected fluxes from the SALT2 model, and this 
mismatch will introduce systematic errors on the measurements. To estimate these errors 
we created a new filter set by applying a random offset to the central wavelength 
$\lambda_{\mathrm{c}}$ of each filter. This offset was drawn from a uniform distribution limited 
to $\pm 2.5\times 10^{-3}\lambda_{\mathrm{c}}$, which is a conservative specification for the J-PAS 
filters \citep{MarinFranch12mn}. 

The SN fluxes were simulated with this new set of filters, and the simulated measurements 
were fitted both with our fiducial filter set (thus introducing the mismatch between assumed and 
actual filters) and with the same set used to simulate the data (which served as a 
systematics--free reference). The best-fitting SALT2 parameters under the two filter sets were 
compared for each individual SN Ia.

\begin{figure}
\includegraphics[width=1\columnwidth]{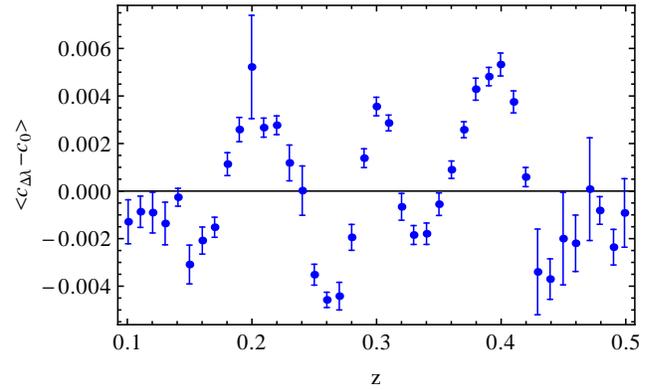}
\caption{Average difference between the recovered values for the SALT2 colour 
parameter when the SNe Ia are fitted using mismatched and correct filter sets, as 
a function of redshift. The use of a synthetic filter set with wrong central wavelengths 
introduces a small oscillatory bias to the measured SALT2 parameters.} 
\label{fig:filter-shift-bias} 
\end{figure}

The mismatch between the true and assumed central wavelengths introduces a redshift dependent bias 
to the SALT2 parameters which frequently presents oscillations, specially for $c$, $m_B$ and $\mu$, 
while $t_0$ is the least affected parameter. These oscillations have a period within the range 
$0.08\la \Delta z\la 0.14$, while their amplitude depends on the average offset applied and their 
phase does not follow any clear relation. Fig. \ref{fig:filter-shift-bias} shows an example 
of this bias for the colour parameter.

The oscillatory characteristic of the bias might be explained by the series 
of peaks and troughs in the SN Ia spectrum that, at its peak luminosity, repeat itself 
approximately every 500 \AA. To understand how these two could be tied, imagine 
that a certain blue filter has an assumed central wavelength $\lambda_0$ and a true central 
wavelength $\lambda_\Delta$, shifted to a smaller value. At a certain redshift $z_1$, $\lambda_0$ 
coincide with a spectral peak while $\lambda_\Delta$ do not. Therefore, we will measure a flux smaller 
than expected for $\lambda_0$ and conclude that the SN is redder than in reality. For a SN Ia at 
a higher redshift $z_2$, when $\lambda_0$ coincides with a spectral trough, the 
measured flux will be higher than expected for $\lambda_0$ and we will conclude that the SN 
is bluer. The cycle repeats at a higher redshift $z_3=z_1+\Delta z$ when another peak appears 
at $\lambda_0$. While the exact $\Delta z$ needed to make two consecutive peaks appear in the same 
filter depends on $z_1$ and $\lambda_0$, it amounts to $\sim 0.12$ for the average filter and 
redshift.

In a real survey the filter central wavelength would vary across the field, 
different observations for each SNe would be dithered and each SNe would be observed in different regions 
of the filter. On one hand this eliminates part of the systematic error as statistical error, which 
in turn gets overwhelmed by the other error sources; this makes our estimates for this systematic 
uncertainty more conservative. On the other hand, the inhomogeneity of the filters and the dithering technique 
result in the galaxy subtraction being performed at slightly different wavelengths, a potentially 
harmful effect, specially if sharp galaxy spectral features are included or excluded by the wavelength 
shift. This might introduce strong random fluctuations in the photometry that are not represented by the 
error estimates. Such effect cannot be modelled in {\sc snana} and will not be analysed here, but it 
should be investigated in the future. 

\begin{table}
\caption{Estimated systematic errors on the SALT2 parameters caused by shifts on the filter 
central wavelengths, each randomly drawn from a uniform distribution within the range 
$\pm0.25\%$ and $\pm0.08\%$ of the filter's central wavelength.}
\label{tab:filter-shift-bias}
\centering
\begin{tabular}{|c|c|c|c|c|c|}
\hline
$\Delta\lambda_{\mathrm{c}}$ range & $\sigma_{m_B}^{\mathrm{sys}}$ & $\sigma_{x_1}^{\mathrm{sys}}$ & 
$\sigma_c^{\mathrm{sys}}$ & $\sigma_{t_0}^{\mathrm{sys}}$ & $\sigma_{\mu}^{\mathrm{sys}}$ \tabularnewline\hline
$\pm0.25\%$               & 0.0053        & 0.041        & 0.0034     & 0.047         & 0.0070        \tabularnewline
$\pm0.08\%$               & 0.0015        & 0.014        & 0.0010     & 0.013         & 0.0021         \tabularnewline
\hline
\end{tabular}
\end{table} 

The resulting $1\sigma$ systematic errors caused by the mismatch between 
assumed and real filter wavelengths are shown in Table \ref{tab:filter-shift-bias}. 
We also present the estimated errors for offsets between $\pm 0.8\times 10^{-3}\lambda_{\mathrm{c}}$, 
roughly the precision one would get by characterising the filters with a spectrophotometer 
and using the measured transmission curves as the synthetic ones. Lastly, we remark that 
mismatched filters could also introduce biases in the host galaxy photo-$z$s which in turn 
could impact the SN parameter measurements. Unfortunately, the effect of mismatched filters 
on the galaxy photo-$z$ is beyond the scope of this paper, although we verified that their 
effect on the SN photo-$z$s leads to systematic uncertainties of 0.001 and 0.0004 for offsets between 
$\pm 2.5\times 10^{-3}\lambda_{\mathrm{c}}$ and $\pm 0.8\times 10^{-3}\lambda_{\mathrm{c}}$, respectively, 
which are much less than the photo-$z$ rms errors of 0.005. We also verified the impact 
of constant photo-$z$ biases in SN Ia parameters in section \ref{sec:host-z-bias}. We emphasise 
that for J-PAS, in particular, filters will be fully characterised so that the central wavelength 
offsets will be smaller than $\pm 0.8\times 10^{-3}\lambda_{\mathrm{c}}$.  Moreover, the galaxy photo-$z$s 
will be computed from a stack of dithered images, further reducing the average offset and 
its influence to negligible levels.

\subsection{Calibration biases}
\label{sec:calibration-bias}

To test the effects of photometry calibration biases on the SN data we applied random 
offsets to each filter zero point, drawing from a Gaussian distribution with 
standard deviation $\sigma_{\mathrm{\Delta ZP}}=0.01$, the same precision expected for 
DES \citep{Bernstein12mn}. The application of a zero point recalibration technique based on
photometric redshift estimations from emission line galaxies \citep{Molino14mn} might make 
this level of bias a conservative estimate. We adopted random offsets as a 
simplifying assumption since the specification of more complex biases would require detailed 
analysis of calibration methods which is beyond the scope of this paper.

As with shifts on the filter central wavelengths, the resulting SALT2 parameter biases 
from zero point offsets are usually redshift dependent. However, no clear common pattern 
could be identified: various realisations of the bias may lead to different general trends, offsets and 
fast variations on SALT2 parameters. Table \ref{tab:zp-bias} presents the average difference 
between SN Ia fits with and without the calibration bias. Their values are of 
the same order of the $\pm 0.8\times 10^{-3}\lambda_{\mathrm{c}}$ shift on 
the filter central wavelengths presented in section \ref{sec:central-wavelength}. The 
systematic uncertainty on the SN photo-$z$ resulting from this calibration bias was 0.0004, 
also comparable to the $\pm 0.8\times 10^{-3}\lambda_{\mathrm{c}}$ offset uncertainty. 

\begin{table}
\caption{Estimated systematic errors on the SALT2 parameters caused by offsets 
on the filter zero points, each randomly drawn from a Gaussian distribution with 
$\sigma_{\mathrm{\Delta ZP}}=0.01$.}
\label{tab:zp-bias}
\centering
\begin{tabular}{|c|c|c|c|c|c|}
\hline
$\sigma_{\mathrm{\Delta ZP}}$ & $\sigma_{m_B}^{\mathrm{sys}}$ & $\sigma_{x_1}^{\mathrm{sys}}$ & 
$\sigma_c^{\mathrm{sys}}$ & $\sigma_{t_0}^{\mathrm{sys}}$ & $\sigma_{\mu}^{\mathrm{sys}}$ \tabularnewline\hline
0.01 & 0.0019        & 0.014        & 0.0012     & 0.015         & 0.0024        \tabularnewline
\hline
\end{tabular}
\end{table} 

\subsection{Photo-$z$ biases}
\label{sec:host-z-bias}

Typical systematic galaxy photo-$z$ biases are about 0.33 of the rms error, 
and using spectroscopy to calibrate the photo-$z$s might help reducing it. To test the effects 
of such bias on SN Ia data we created four different simulations, each one with a 
constant offset on the host galaxy photo-$z$s: $\pm0.001(1+z)$ and $\pm0.002(1+z)$.

The average effect of a photo-$z$ bias on SN colour is simple: if the redshift estimate is higher 
than its true value, the SN image will seem bluer than expected for that redshift and the 
inferred colour will be smaller than its true value. If the redshift estimate is lower than 
the true value, the SN will seem redder. On the other hand, the 
effect on the $x_1$ estimate is more complex since two distinct redshift dependent 
effects compete: light curve time dilation and spectral shift in wavelength. Given that 
light curves are stretched by redshift (time intervals are longer at higher redshifts -- 
see Eq. \ref{eq:salt2-flux}), assuming a smaller redshift for the SN will lead to 
a larger $x_1$ estimate since it will have to compensate for the unaccounted extra 
bit of time dilation. In opposition, light curves on the bluer part of the SN spectrum 
are usually narrower than redder light curves (see Fig \ref{fig:spectral-surface}). 
Therefore, a smaller redshift estimate will make one take a bluer light curve for a red one, 
pushing the $x_1$ estimate to smaller values. The resulting $x_1$ bias from these two 
effects depends on the redshift and filter set, and in our particular case, 
the spectral shift effect seems to be slightly larger for $z\la 0.3$ while time dilation 
dominates at $z\ga 0.3$.

The apparent magnitude $m_B$ is also affected by competing effects: at lower redshifts, 
the SN spectrum is more compact in wavelength space (there would be more 
photons per unit wavelength), so underestimations of $z$ make the SN look fainter. 
However, the SN rest-frame spectrum peaks at $\sim4000$ \AA\, -- almost 
outside our filter set wavelength range -- so underestimations of $z$ leads to the 
wrong conclusion that one is measuring fainter regions of the SN spectrum, 
therefore increasing the inferred luminosity.  

Finally, the distance modulus $\mu$ is affected by the biases in $m_B$, $x_1$ 
and $c$ and by a combination of Malmquist bias and misguided estimates of the 
nuisance parameters $\alpha$ and $\beta$: at higher redshifts, our survey 
preferentially detects more luminous SNe Ia, which would lead to underestimations 
of the luminosity distance. Since these SNe tend to be bluer, the term $-\beta c$
in Eq. \ref{eq:mB-def} partially corrects for this effect. However, the 
biases in $x_1$ and $c$ induce slightly off $\alpha$ and $\beta$ values which 
will under or over-correct distance measurements. We remind that all the processes presented 
here describe average effects on SNe Ia data. The effects of a photo-$z$ bias on each 
individual SN is much harder to predict or describe. Table \ref{tab:z-bias} presents 
the estimates for the systematic errors on SALT2 parameters due to photo-$z$ biases 
of order $0.001(1+z)$ and $0.002(1+z)$.

\begin{table}
\caption{Estimated systematic errors on the SALT2 parameters caused by $0.001(1+z)$ 
and $0.002(1+z)$ systematic biases in the photo-$z$s.}
\label{tab:z-bias}
\centering
\begin{tabular}{|c|c|c|c|c|c|}
\hline
$\Delta z/(1+z)$ & $\sigma_{m_B}^{\mathrm{sys}}$ & $\sigma_{x_1}^{\mathrm{sys}}$ & 
$\sigma_c^{\mathrm{sys}}$ & $\sigma_{t_0}^{\mathrm{sys}}$ & $\sigma_{\mu}^{\mathrm{sys}}$ \tabularnewline\hline
$\pm 0.002$               & 0.0040        & 0.031        & 0.0037     & 0.062         & 0.0096        \tabularnewline
$\pm 0.001$               & 0.0018        & 0.012        & 0.0019     & 0.031         & 0.0059         \tabularnewline
\hline
\end{tabular}
\end{table}

In case of a photo-$z$ bias in which the offsets $\Delta z_i$ applied to each 
SN $i$ have $\langle\Delta z_i\rangle \neq 0$ (such as the constant bias we simulated), 
$\mu$ will get an extra constant offset, roughly of order 
$\mu_0(\bar{z}+\langle\Delta z_i\rangle)-\mu_0(\bar{z})$, due to a bias on the nuisance parameter $M$. 
Here, $\mu_0$ is the fiducial distance modulus used to estimate the nuisance parameters 
(see section \ref{sec:typing-fitting-process}) and $\bar{z}$ is the average 
redshift of the survey. Since constant offsets in $\mu$ are irrelevant for 
many applications, these are not included in Table \ref{tab:z-bias}.

\section{Summary and conclusions}
\label{sec:conclusions}

We used the {\sc snana} software package and the SALT2 model to simulate 
the SN Ia data that a narrow band survey could obtain. We adopted J-PAS 
as our survey model, which is going to image 8500 $\mathrm{deg^2}$ of the 
sky in 54 narrow band ($\sim 100$ \AA) and five broad band filters (see Fig. 
\ref{fig:filter-transmission} for the transmission curves of the unique 
filters) and that can reach a galaxy photo-$z$ precision of $0.005(1+z)$ 
\citep{Benitez14mn}. The observing strategy we assumed is called 
2+(1+1). Each field would be imaged four times in each one of the 56 filters:  
twice during the same night in order to measure the flux from host galaxies 
and twice in different nights (spaced by $\sim$ one month) in order to find 
SNe and measure their light curves. In each night, 14 different filters are 
imaged, and the gap between different sets of filters is $\sim$ one week 
(Fig. \ref{fig:cadence-M01} shows a graphical representation of this observation 
schedule).

First, we showed in section \ref{sec:results} that such an SN survey 
is indeed possible and can yield precise measurements of spectral features 
(see Figs. \ref{fig:low-z-spec} and \ref{fig:med-z-spec}), light-curve parameters 
(Figs. \ref{fig:bias-rms-mB}--\ref{fig:bias-rms-mu} and Table \ref{tab:avg-rms-M01}), 
SN photo-$z$s (Fig. \ref{fig:sn-photo-z}) and can achieve low contamination fractions 
by CC-SNe (see Table \ref{tab:typing-results}), all without any spectroscopic 
follow-up. This might be surprising since each light curve only has two measurement 
points. To understand this result, one should bear in mind that there are 
56 light curves (making a total of 112 observations) and they all are being 
described by the same 5 parameters. This is even clearer if one thinks in 
terms of constraining a spectral surface $f_{\lambda}(\lambda_{\mathrm{obs}},t_{\mathrm{obs}})$ 
instead of light curves (see Fig. \ref{fig:spectral-surface}). We also 
studied potential systematics (section \ref{sec:systematics}) -- with 
photo-$z$ biases being the most relevant ones -- and showed that they are 
small (less than 0.10 of the rms). The systematic uncertainties on 
the SNe Ia parameters are also more than 10 times smaller than known differences 
between SNe Ia in different environments \citep[e.g.][]{Xavier13mn}, so our fiducial survey 
still has room for the discovery of smaller, unknown potential differences. 

On top of the precision attainable, a telescope with a large field of view may lead 
to massive samples (approximately 500 SNe Ia and 90 CC-SNe every two months 
-- see Table \ref{tab:sample-sizes}) that can be used in the study of rates, 
spectral feature relations, dust extinction and intrinsic colour variations and 
correlations between SN and environment properties. Besides the increase in 
sample size, most of these topics can also benefit from the higher spectral 
resolution when compared to broad band photometry.  

We have also shown that SN narrow band observations can still be optimised 
for better SALT2 parameters constraints 
by (a) better distributing the observations over $f_{\lambda}(\lambda_{\mathrm{obs}},t_{\mathrm{obs}})$ 
(although one might loose the ability to identify spectral features -- 
see section \ref{sec:dispersed-obs}); and by (b) selecting a smaller set of 
filters that cover the relevant parts of the SN spectra (section 
\ref{sec:no-red-filters}). In the last case the sample size and 
redshift depth are also increased. Other potential strategies for optimising the survey -- 
increasing cadence at expense of exposure time (section \ref{sec:big-cadence}) 
and transferring CCD readout time to exposure time (section \ref{sec:overhead-time}) 
-- proved to be unworthy. Another promising optimising strategy that should be analysed 
in the future is the use of slightly broader filters (up to 200 \AA\, wide) that 
may increase the SNR while maintaining enough spectral resolution to detect SN features. 
For J-PAS in particular, these optimising strategies probably cannot be implemented 
given its other science goals and technical details. For instance, given that its main goal 
is to measure photo-$z$s of high redshift galaxies, the infrared filters 
cannot be put aside to free integration time for the bluer filters.   

On the downside, a narrow band survey is bound to be a low or intermediate 
redshift survey since very long exposure times (or very large telescopes) 
would be needed to substantially increase its depth. Our fiducial survey has an 
average redshift of $\sim0.25$ and reaches a maximum $z\sim0.5$ (see Fig. 
\ref{fig:snia-z-hist}), which is a lot less than ongoing SN surveys 
like DES. Therefore, it may not be competitive to constrain cosmological 
parameters on its own. However, it still can be very valuable for cosmology 
by providing better understanding in the fields mentioned above, which 
enter in cosmological analysis as systematic uncertainties and better 
standardisation methods for SN Ia luminosity.

Although the results presented here are dependent on the adopted specifications
they may serve as a guide for other instruments and observing strategies. 
However, it is important to keep in mind that some characteristics 
are crucial for the survey's performance: a gap of at least one month should 
be provided between the template and the search observations; 
a reasonably wide wavelength range (e.g. 4000--6500 \AA ) 
must be probed in order to provide good colour information; a minimum of four 
search epochs should be available, even if in different filters; and the time interval 
$\Delta t_{\mathrm{s}}$ between different search epochs should be in the range 
$2\la \Delta t_{\mathrm{s}}\la 15$ days. In summary, for the studies mentioned 
above -- which require large SN samples -- a wide-field narrowband survey is 
likely to be the optimal tool since its efficiency surpasses that of broadband 
surveys backed up by spectroscopy and its data quality is greater than 
pure broadband surveys.

\section{Acknowledgements}
\label{sec:acknowledgements}

The authors would like to thank Richard Kessler for expanding the 
{\sc snana} capabilities to a large number of filters, for the 
software support and helpful discussions. 
This work was financially supported by CAPES (BEX 6796/10-9) and FAPESP Brazilian 
funding agencies and has made use of the computing facilities of 
the Laboratory of Astroinformatics (IAG/USP, NAT/Unicsul), whose 
purchase was made possible by the Brazilian agency FAPESP 
(grant 2009/54006-4) and the INCT-A. The Brazilian contribution 
and participation in J-PAS was partially supported by FAPESP 
(EMU: 2009/54162-6). BBS would like to acknowledge financial 
support from the Brazilian funding agency CAPES, grant number 
PNPD-2940/2011.

\bibliographystyle{mn2e}
\bibliography{main}

\appendix

\section{Photometry toy model}
\label{sec:snr-model}

As a guide for the expected outcomes and relationships between a survey's design and the 
resulting flux signals and errors, we developed simplified analytical formulae that 
reproduce the main characteristics and dependencies of the survey's photometry. The 
reader should keep in mind that the work presented in the previous sections involve 
realistic and complex simulations that are capable of uncovering various effects not 
included in this toy model. However, these formulae are useful for pointing out possible 
options for optimising photometry and for understanding the outcomes of the simulations. 

The total CCD counts $C$ is related to the source's apparent magnitude $m$ in the AB system by:

\begin{equation}
C=10^{-0.4[m-ZP^{\mathrm{(\Delta t)}}]},
\label{eq:count-mag-relation}
\end{equation}
where the zero point $ZP^{\mathrm{(\Delta t)}}$ is given by Eq. \ref{eq:zero-point}, which we approximate here to:

\begin{equation}
ZP^{\mathrm{(\Delta t)}}\simeq 2.5\log_{10} \left[ \frac{\pi D^2\Delta t \bar{T}\Delta\lambda}{4h\lambda_{\mathrm{c}}} 
\mathrm{\frac{erg}{cm^2}}\right] - 48.6.
\label{eq:zero-point-aprox}
\end{equation}
In the equation above, $\bar{T}$, $\Delta\lambda$ and $\lambda_{\mathrm{c}}$ are the filter's average 
filter transmission, bandwidth and central wavelength, respectively. 

The relation between the apparent magnitude $m$ and the source's absolute magnitude $M$ is not 
so straightforward. First of all, the photons arriving at the detector with a certain wavelength 
were emitted from a bluer part of the source's spectrum and then redshifted by a factor of $(1+z)$ 
due to the cosmic expansion. Thus, the observed magnitude in a specific filter relates to different 
parts of the source's spectrum depending on the redshift $z$. This effect may be dealt with the 
so-called K-correction, but we ignore it in this toy model by assuming a fixed absolute magnitude 
for all wavelengths:

\begin{equation}
m=M+\mu (z)= M + 5\log_{10}\left[\frac{d_L(z)}{\mathrm{Mpc}}\right] + 25,
\label{eq:mag-mu-aprox}
\end{equation}
where $\mu$ is the distance modulus and $d_L$ is the luminosity distance. 
The result of this approximation depends on the particular filter used and 
on the source's spectrum. For the average J-PAS filter and a SN Ia at peak luminosity, it amounts 
to an underestimation of the CCD counts at high redshifts (which can reach a factor of 1.4 at $z\sim 0.5$)
since the spectra peaks at 4000 \AA\, while most J-PAS filters probe higher wavelengths.

Another effect ignored in Eq. \ref{eq:mag-mu-aprox} is the Malmquist bias: true supernovae present 
a scatter in $M$, and a magnitude-limited survey will preferentially detect brighter objects 
at higher distances. Therefore, the constant $M$ approximation further underestimates the CCD counts 
at high redshifts. 

Lastly, we don't account, in this toy model, for the Galactic extinction, which would reduce the measured 
fluxes according to the filter wavelength and to the SN angular coordinates. For our fiducial survey, 
this approximation would result in an overestimation of the CCD counts by $\sim 15$ per cent. However, 
this effect can roughly be accounted for by an increase on $M$.

To sum up, the approximation described by Eq. \ref{eq:mag-mu-aprox} results in an underestimation of 
CCD counts at high redshifts, where our toy model serves, therefore, as a conservative estimate. 
Nevertheless, it can still be used to understand the general effects of various survey characteristics 
and error sources on the flux measurements.

We considered four kinds of errors in this toy model, separated according to their dependence on 
the survey's parameters and on the source's apparent magnitude: the Poisson noise from the SN 
$\sigma_{\mathrm{SN}}$ and from its host galaxy $\sigma_{\mathrm{g}}$; a multiplicative error 
$\sigma_{ZP}$, given in mag; the CCD readout noise $\bar{\sigma}_{\mathrm{r}}$; and the sky noise 
$\bar{\sigma}_{\mathrm{sky}}$. Assuming there is a set of $N$ exposures of the same field 
which can be averaged into a template for host galaxy subtraction, we can estimate the final error in 
an SN PSF photometry as:

\begin{equation}
\sigma_{\mathrm{C}}=\sqrt{\sigma_{\mathrm{SN}}^2+\left( 1+\frac{1}{N}\right)
  \left(\sigma_{\mathrm{f}}^2+\sigma_{\mathrm{g}}^2+\bar{\sigma}_{\mathrm{r}}^2+\bar{\sigma}_{\mathrm{sky}}^2\right)}, 
\label{eq:total-error}
\end{equation}
where $\sigma_{\mathrm{f}}=[\partial C/\partial ZP^{\mathrm{(\Delta t)}}]\sigma_{ZP}=0.921C\sigma_{ZP}$. 
The Poisson noises $\sigma_{\mathrm{SN}}$ and $\sigma_{\mathrm{g}}$ are simply the square root of Eq. 
\ref{eq:count-mag-relation}.
Finally, $\bar{\sigma}_{\mathrm{r}}$ and $\bar{\sigma}_{\mathrm{sky}}$ are 
related to the errors per pixel, $\sigma_{\mathrm{r}}$ and $\sigma_{\mathrm{sky}}$, by the formula:

\begin{equation}
\bar{\sigma}_x=\sqrt{4\pi\sigma_{\mathrm{PSF}}^2}\sigma_x,
\label{eq:err-pixel-to-psf}
\end{equation}
where $\sigma_{\mathrm{PSF}}$ is the PSF (assumed Gaussian) radius in pixels. The sky noise per pixel 
is given by Eq. \ref{eq:sky-sig}, which can be simplified for narrow band filter in order to 
get rid of the exact shape of the filter bandwidth:

\begin{equation}
\sigma_{\mathrm{sky}}^2=P^210^{-0.4[m_{\mathrm{sky}}-ZP^{\mathrm{(\Delta t)}}]},
\label{eq:sky-sig-simple}
\end{equation}
where $P$ is the pixel angular size in arcsec and $m_{\mathrm{sky}}$ is the sky magnitude per 
$\mathrm{arcsec}^2$. Although Eq. \ref{eq:total-error} is valid for an individual data point, 
one should keep in mind that if two or more subtracted images share common templates their 
errors will be correlated, which turn the analysis of the set of these images into a more complex 
subject. Average values that can be used with these formulae are listed in Table 
\ref{tab:survey-facts} and the J-PAS filter characteristics are presented in Fig. 
\ref{fig:filter-transmission}. Coarse values for absolute magnitudes are 
$M_{\mathrm{SN}}=M_{\mathrm{g}}=-18.4$ and $m_{\mathrm{sky}}=20.2$.  

Since {\sc snana v10\_29} assumes very deep templates ($N\rightarrow\infty$), we emulated 
shallow templates in the simulations by inflating the inputs used to generate each noise term: 
$\sigma_{ZP}$ was overestimated by a factor of $\sqrt{1+1/N}$, $\sigma_{\mathrm{sky}}$ was replaced 
according to Eq. \ref{eq:inflated-sky} and the galaxy magnitudes $m_{\mathrm{g}}$ used to compute 
$\sigma_{\mathrm{g}}$ were substituted by

\begin{equation}
m_{\mathrm{g}}'=m_{\mathrm{g}}-2.5\log_{10}\left( 1+\frac{1}{N}\right).
\label{eq:inflated-gal-mags}
\end{equation}

\end{document}